\documentclass[12pt,twoside]{article}   

\usepackage{fancyhdr}		

\usepackage[section]{placeins}   %

\usepackage{graphicx}

\graphicspath{{PaperFigures/}}

\renewcommand{\thesection}{{\sf \Roman{section}}.}
\renewcommand{\thesubsection}{\thesection*{\sf \Alph{subsection}}.}
\renewcommand{\thesubsubsection}{\thesubsection*{\sf \arabic{subsubsection}}.}

\newcommand{\mfig}[1]{\marginpar{{\sf Fig~\ref{#1} }}}
\newcommand{\mtab}[1]{\marginpar{{\sf Table~\ref{#1} }}}

\renewcommand{\refname}{}       %

\usepackage{hyperref}
\hypersetup{ colorlinks,
    citecolor=blue,
    filecolor=blue,
    linkcolor=blue,
    urlcolor=blue
}

\usepackage{xcolor}

\definecolor{gray}{rgb}{0.6,0.6,0.6}
\definecolor{red}{rgb}{0.85,0,0}
\definecolor{green}{rgb}{0,0.85,0}
\definecolor{blue}{rgb}{0,0,0.85}
\definecolor{beige}{rgb}{0.92,0.87,0.78}
\usepackage[all]{hypcap}    

\usepackage{float}
\usepackage{subfiles}
\usepackage{subcaption}
\usepackage{makecell}
\usepackage{nameref}
\usepackage{booktabs}
\usepackage{amsmath}
\usepackage{geometry}
\usepackage{setspace}
\usepackage{nameref} 

\usepackage{authblk}

\usepackage{xr}
\newcommand{\NPatients}{137}

\newcommand\Lof{L\"ofstedt}

\newcommand{\ie}{{\it i.e.},}
\newcommand{\eg}{{\it e.g.},}
\newcommand{\etal}{{\it et al}}

\newcommand\LowerQuantProstate{18}
\newcommand\UpperQuantProstate{2500}

\newcommand{\DDTG}{\%$\Delta_{f}$}
\newcommand{\DDninty}{\%$\Delta_{D90}$}

\begin{document}

\title{\bfseries \Large Haralick texture feature analysis for Monte Carlo dose distributions of permanent implant prostate brachytherapy}



\author[1]{Iymad R. Mansour}
\author[2]{Nelson Miksys}
\author[3]{Luc Beaulieu}
\author[4]{\'Eric Vigneault}
\author[1,*]{Rowan M. Thomson}

\affil[1]{Carleton Laboratory for Radiotherapy Physics, Physics Department, Carleton University,
             1125 Colonel By Dr,
             Ottawa,
             K1S 5B6,
             Ontario,
             Canada}
             
\affil[2]{Ackerman Cancer Center, 
10881 San Jose Blvd,
Jacksonville,
32223,
Florida, 
USA}

\affil[3]{D\'epartement de Physique et Centre de recherche sur le cancer, Service de physique m\'edicale et radioprotection et Centre de recherche du CHU de Qu\'ebec, Universit\'e Laval,
             9 Rue McMahon,
             Qu\'ebec City,
             G1R 3S3,
             Qu\'ebec,
             Canada}

\affil[4]{Centre de recherche sur le cancer, D\'epartement de Radio-Oncologie et Centre de recherche du CHU de Qu\'ebec, Universit\'e Laval,
             9 Rue McMahon,
             Qu\'ebec City,
             G1R 3S3,
             Qu\'ebec,
             Canada}

\affil[*]{rthomson@physics.carleton.ca}

\newgeometry{top=0in,bottom=1in,right=1in,left=1in}
\thispagestyle{fancy}
\begin{titlepage}
\maketitle


\begin{abstract}

\noindent 
{\bf Purpose:} Demonstrate quantitative characterization of 3D patient-specific absorbed dose distributions using Haralick texture analysis, and interpret measures in terms of underlying physics and radiation dosimetry. 

\noindent
{\bf Methods:} Retrospective analysis is performed for 137 patients who underwent permanent implant prostate brachytherapy using two simulation conditions: ``TG186'' (realistic tissues including 0-3.8\% intraprostatic calcifications; interseed attenuation) and ``TG43'' (water-model; no interseed attenuation). Five Haralick features (homogeneity, contrast, correlation, local homogeneity, entropy) are calculated using the original Haralick formalism, and a modified approach designed to reduce grey-level quantization sensitivity. Trends in textural features are compared to clinical dosimetric measures (D90; minimum absorbed dose to the hottest 90\% of a volume) and changes in patient target volume \% intraprostatic calcifications by volume (\%IC). 

\noindent
{\bf Results:} Both original and modified measures quantify the spatial differences in absorbed dose distributions. Strong correlations between differences in textural measures calculated under TG43 and TG186 conditions and \%IC are observed for all measures. For example, differences between measures of contrast and correlation increase and decrease respectively as patients with higher levels of \%IC are evaluated, reflecting the large differences across adjacent voxels (higher absorbed dose in voxels with calcification) when calculated under TG186 conditions. Conversely, the D90 metric is relatively weakly correlated with textural measures, as it generally does not characterize the spatial distribution of absorbed dose.  

\noindent
{\bf Conclusion:} patient-specific 3D dose distributions may be quantified using Haralick analysis, and trends may be interpreted in terms of fundamental physics. Promising future directions include investigations of novel treatment modalities and clinical outcomes. 

\vspace{1em}

\end{abstract}
\end{titlepage}

\newgeometry{top=1.0in,bottom=1in,right=1in,left=1in}


\newpage

\setlength{\baselineskip}{0.7cm}      

\pagenumbering{arabic}
\setcounter{page}{1}
\pagestyle{fancy} 

\fancyhf{} 
\fancyfoot[R]{\thepage}

\fancypagestyle{plain}{%
    \renewcommand{\headrulewidth}{0pt}%
    \fancyhf{}%
    \fancyfoot[R]{\thepage}%
}


\section*{Introduction}

In the field of medical physics, absorbed dose \cite{ICRU85} is of primary interest to evaluate radiation therapy treatments and inform advancements within the field. In recent years, there has been a general shift towards the consideration of increasingly realistic patient specific data to improve accuracy of absorbed dose calculations \cite{TG105,TG106,TG185,TG186}, in part, to facilitate improved understanding of clinical outcomes from radiation treatments. For example, heterogeneities in absorbed dose distribution calculations, resulting from heterogeneities in material compositions of patient specific models, have been identified to be correlated with patient outcomes \cite{Ev19}. Despite advancements in calculations of absorbed dose and demonstrated importance of the 3D spatial distribution of absorbed doses, the  approaches used to investigate dosimetric data often rely on rudimentary statistics (\eg{} mean or standard deviation). Commonly used clinical metrics calculated from a Dose Volume Histogram (DVH) such as D90  or Dose Homogeneity Index (DHI \cite{Mi17}) inherently ignore the spatial distribution of data. The purpose of this work is to address the shortcomings of these classical dose metrics by presenting, and examining, analysis approaches that are sensitive to the spatial distribution of a patient’s dose distribution.



Sophisticated methods exist for the examination and quantification of the spatial distribution of data via the field of texture analysis \cite{Ma04, MM98}. Texture analysis has been used extensively across many scientific disciplines, \eg{} image processing \cite{ST99, Ca01} or computer vision \cite{Dp12}. Recently, texture analysis has been applied to medical physics and bioinformatics (\eg{} target volume contour evaluation \cite{Zh19} or Raman spectroscopic mapping of radiation response \cite{Vi19}), and is providing the impetus for development of novel subfields within medical physics. The fields of radiomics \cite{Vm15}, and dosiomics \cite{Rl18} use texture analysis to investigate medical images and dosimetric data respectively; primarily for the prediction of a subset of clinical treatment outcomes (\eg{} radiation pneumonitis \cite{Li19, Ta21}). Although the capability of texture analysis is well demonstrated by its use within such diverse applications, extracted textural measures are seldom applied for quantitative analysis, where calculated measures are interpreted and used to characterize datasets. 

Recently, the first application of texture analysis to dosimetric data for the purpose of quantitative characterization was successfully demonstrated in a proof-of-concept study considering simplified scenarios. \cite{MT22} used Haralick features \cite{Ha73} to illustrate the potential of texture analysis for the characterization of dosimetric data across varying length scales. Haralick analysis is a particular form of textural analysis which was initially proposed as a method to quantify the relation of neighbouring pixels within images, but has since been implemented widely across numerous fields for the analysis of diverse types of data, due in large part to its simplicity and intuitive interpretations. By considering dosimetric data (\eg{} absorbed dose deposited within a voxel) as analogous to pixel intensity, \cite{MT22} demonstrated the application of Haralick analysis to dosimetric data calculated using Monte Carlo (MC) simulations and related trends in extracted textural measures to radiation physics and dosimetry.

Haralick analysis uses a grey level co-occurrence matrix (GLCM) to quantify textural measures. Computation of a GLCM requires that all data being analyzed first undergo a ``quantization'' process \cite{Ha73}. Typically, the quantization process involves mapping intensity values within a distribution (\eg{} pixel intensity within an image, or magnitude of absorbed dose deposited within a voxel) to a finite number of bins (typically referred to as ``grey levels'', with a maximum number of grey levels: $N$). There exists an established dependence of the calculated textural measures to quantization \cite{Tl19, Bp17, Ca02, Dg17}. Despite this, there is no standardized approach to quantization for Haralick texture analysis \cite{Rl15}. For example, some studies have investigated both different mapping techniques and the sensitivity of extracted textural measures to the number of grey levels considered; in general, applications within the literature typically adopt a data-specific approach to quantization \cite{Gw12, ST99, Ca02, Dg17,Va17}. In some applications, little information is provided to motivate or justify approaches to quantization, and this may represent a limitation of current applications of texture analysis. 

To address the sensitivity to quantization, \Lof{} \etal{} \cite{Tl19} developed a modified set of Haralick textural measures intended to be asymptotically invariant to the choice of $N$. \Lof{} \etal{} \cite{Tl19} developed these features by modifying the original Haralick measures, with the intention of creating textural measures which are less sensitive to quantization but that preserve most of the original measures interpretations. Although these features represent an important advancement in texture analysis, and are demonstrated to be  less sensitive to quantization, recent work has demonstrated the continued importance of considering well motivated or data-specific quantization approaches \cite{Ga18, MT23}. 


The work presented has two novel objectives extending applications of texture analysis for quantitative characterization of dosimetric data which have been presented in the literature \cite{MT22, MT23}:  explore the first application of textural features to patient specific 3D dosimetric data for quantitative analysis, and compare different approaches of texture analysis by examining both original and modified Haralick textural features in this context. For the present work we consider a single treatment type to explore the Haralick analysis approach with clinical patient data, focusing on $^{125}$I permanent implant prostate brachytherapy (PIPB) patient data. PIPB offers dosimetric data which are particularly interesting to explore using texture analysis, some examples include: highly non-uniform dose distributions \cite{Na00}, sharp dose gradients due to intraprostatic calcifications \cite{Mi17}, and cold spots due to interseed attenuation \cite{Ma14, Sa13}. Considering those PIPB 3D dose distributions, we calculate original and modified textural measures and investigate sensitivity to anatomical differences between patients (intraprostatic calcification as a percent of the total prostate volume; \%IC), traditional clinical metrics of $D90$ and Dose Homogeneity Index (DHI), and quantization. Original and modified textural measures are compared based on their sensitivity to quantization, and their underlying mathematical differences.

\section*{Methods}

\subsection*{Patient specific Monte Carlo absorbed dose distributions}\label{meth:mc}

MC simulations for \NPatients{} patients who underwent $^{125}$I (Nucletron SelectSeed, ProstaSeed 125SL, Draximage LS-1, and OncoSeed 6711) permanent implant prostate brachytherapy (PIPB) were previously carried out using the EGSnrc \cite{egsnrc} user-code BrachyDose \cite{Ta07} and a brief summary of the calculations is provided (for full details see  \cite{Mi17}). Virtual patient models were derived from post implant Computed Tomography (CT) images with physician-drawn target, urethra, rectum, and bladder contours. CT images were acquired using a helical scan with 2 mm slices. Virtual models used for calculations of absorbed dose preserved the voxel dimensionality from the acquired CT images: voxel volume resolutions were uniform for each patient, but vary across the patient cohort considered (0.21-0.29 mm$^3$). Within the virtual models, each voxel's mass density and tissue elemental composition were determined from CT number (after reduction of metallic artefacts \cite{Mi15}) and voxel location relative to contours. For calcifications specifically, voxels with a mass density greater than 1.27 g/cm$^3$ located within the prostate volume were assigned to calcification. Voxels assigned to calcification were all confirmed by visual inspection. 

The intraprostatic calcification as a percent of the total prostate volume (\%IC) varies across the patient population (0-3.8\%) with 67 of the \NPatients{} patients analyzed having some level of intraprostatic calcification. Across the patients considered, the number of sources varies from 31 to 83, and air kerma strength varies from 0.43-0.78 U.
Simulations initialized 10$^9$ particles resulting in maximum statistical uncertainties of less than 0.5\% to the target, rectum, bladder, and urethra-contoured volumes. Calculations of absorbed dose using two conditions are considered, and have the following shorthand notation:

\begin{enumerate}
\item TG43: simulating the TG43 \cite{TG43U1} formalism with each voxel’s elemental composition, mass density, and scatter material assigned to that of 0.998 g/cm$^{3}$ water and interseed attenuation not modelled; and
\item TG186: model-based dose calculations with tissue-specific elemental compositions and mass densities modelled in each voxel, interseed attenuation modelled \cite{TG186}.
\end{enumerate}

\subsection*{Haralick analysis}\label{meth:haralick}

\begin{figure}[H]
	\centering
	\includegraphics[width=0.8\linewidth]{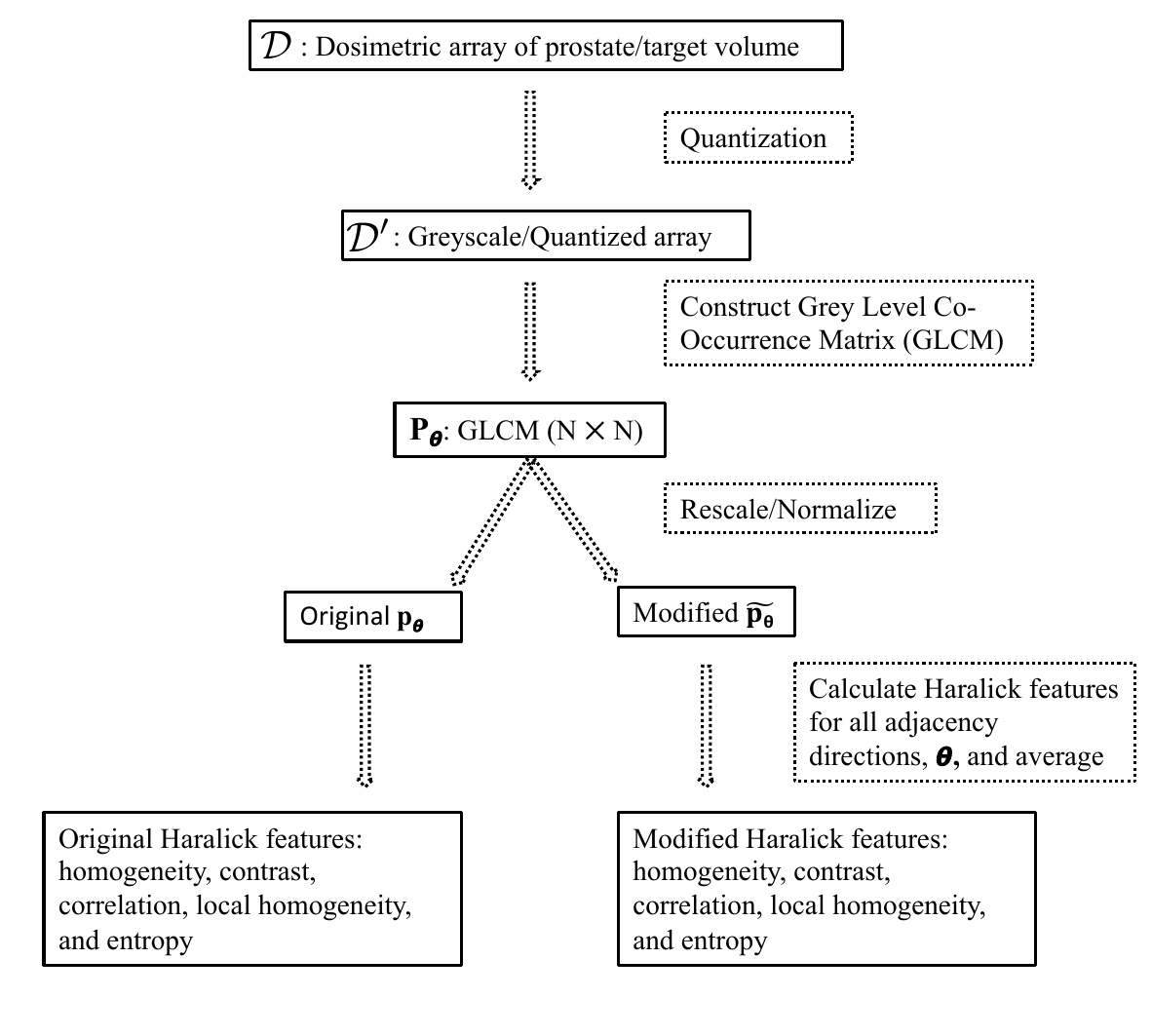}
\caption[Overview of the framework to generate dose distributions, process, and extract directionally invariant original and modified textural features from patient specific PIPB absorbed dose distributions]{Overview of the framework to generate dose distributions, process, and extract directionally invariant original and modified textural features.}
	\label{fig:flowChartGrouped}
\end{figure}

The flowchart in figure~\ref{fig:flowChartGrouped}\mfig{fig:flowChartGrouped} provides an overview of the calculation of original and modified Haralick measures from absorbed dose distributions. 

Absorbed dose distributions within the target prostate volume ($\mathcal{D}$) are isolated from the larger 3D patient specific absorbed dose distribution; all calculations of Haralick measures investigate absorbed dose deposited exclusively within the target volume. Distributions of absorbed dose are quantized ($\mathcal{D'}$) by mapping each value of absorbed dose to an integer value representing quantization levels. A linear quantization scheme is applied to all absorbed dose distributions. The quantization approach implemented herein uses the \texttt{digitize()} function (from Python \cite{python} package \texttt{numpy} \cite{numpy}) with the following bin delimiters,

\begin{equation}\label{eq:mapping}
	b_{i} = D_{\mathrm{1}} + \frac{i\cdot (D_{\mathrm{2}}- D_{\mathrm{1}})}{N_b}, \quad i= 0, 1, 2, ..., N_{b},
\end{equation}

\noindent
where $N_{b}$ is the maximum number of possible quantization levels, $D_{1}= \LowerQuantProstate{}$~Gy and $D_{2}=\UpperQuantProstate{}$~Gy  corresponding to the smallest and largest magnitude of absorbed dose, respectively (considering all datasets, excluding any voxels directly intersecting with seeds and the remaining hottest 0.1\% of voxels; selected to limit the effect of very high values of absorbed dose directly adjacent to seeds). All voxels with dose higher than $D_{2}$ are assigned to $N_b$. The integer $N_b$ determines the width of the quantization levels  as $\Delta D = (D_2-D_1)/N_b$. Considering an upper limit of \UpperQuantProstate{} Gy results in all quantized distributions having maximum number of grey levels ($N$) equal to $N_b$. The analyses of Haralick measures (Results sections: ~\nameref{sec:res:rawFeatures} and \nameref{sec:res:patientCohort}) present Haralick features determined with $N_{b}= 1000$. The sensitivity of extracted textural measures to $N_{b}$ is investigated in Results section ~\nameref{sec:res:compareApproaches}; Table~\ref{tab:deltaD}\mtab{tab:deltaD} presents the different values of $\Delta D$, the quantization level ``width'', alongside corresponding values of $N_b$, considered.

\begin{table}[H]
\vspace{1em}
  \centering
  \caption[The maximum number of quantization levels ($N_{b}$), and corresponding quantization level width ($\Delta D$), considered for the investigation of textural measures to quantization for patient specific absorbed dose distributions.]{The maximum number of quantization levels ($N_{b}$), and corresponding quantization level width ($\Delta D$), considered for the investigation of textural measures to quantization (Results section ~\nameref{sec:res:compareApproaches}).}
\begin{singlespace*}
    \begin{tabular}{cc} \toprule
    $N_{b}$ & $\Delta D$ / Gy \\ \midrule
    50   & 50 \\
    100   & 25 \\
    200   & 13 \\
    250   & 10 \\        
	500   & 5.0 \\ 
    1000   & 2.5 \\
    3000   & 0.84 \\
    5000   & 0.50 \\
    7500   & 0.34 \\        
	10000   & 0.25 \\\bottomrule
    \end{tabular}%
    \end{singlespace*}
  \label{tab:deltaD}%
\end{table}%

The function \texttt{cooccurence} within the package \texttt{mahotas} \cite{Cl13} available for Python \cite{python} is used for the calculation of all grey level co-occurrence matrices (GLCM), $\textbf{P}_{\theta}$. Each element $P_{\theta}(i, j)$ in this $N \times N$ matrix, where $N$ represents the largest value in the quantized array $\mathcal{D'}$ ($N \leq N_b$), counts the number of times an adjacency pair, adjacent elements with values $i$ and $j$, appears in the quantized distribution $\mathcal{D'}$ within a displacement of one voxel and for a particular adjacency direction $\theta$ within the 3-dimensional rectilinear absorbed dose distributions analysed within this work. To increase efficiency, symmetric GLCMs (which are semi-directional invariant and represent the bi-directional relationship between adjacent elements \cite{Tl19}) are used for the computation of GLCMs, and GLCMs representing all 26 possible adjacency directions within the 3-dimensional absorbed dose distribution. Calculated GLCMs are used to both calculate original and modified Haralick measures (Results sections: \nameref{sec:methods:OGHaralick} and \nameref{sec:methods:InvHaralick} respectively).

\subsubsection*{Original Haralick measures}\label{sec:methods:OGHaralick}

The \texttt{haralick()} function available within the \texttt{mahotas} \cite{Cl13} package for the Python programming language \cite{python} uses the calculated GLCMs to compute normalized GLCM's for each adjacency directions ($\textbf{p}_{\theta}$, by dividing each element by the sum, $\sum_{i,j=1}^{N}P_{\theta}(i,j)$), and all 14 textural measures proposed in the original work by Haralick. The amount of time required to compute textural measures is dependent on the magnitude of $N_{b}$ considered.  As an example, calculations using an $N_{b}$ of 50, 1000, and 10000 took an average of 0.22 $\pm$ 0.7, 0.90 $\pm$ 0.14, and 105 $\pm$ 32 seconds to complete on a single Intel Xeon processor (E5-2683). Each measure is averaged over all adjacency directions to compute directionally invariant textural measures. The computation of directionally invariant measures is an approach which is consistent with the literature \cite{Vi19,Bp17,Vm15}. Due to a high level of correlation between calculated measures, this work follows the approach by Vrbik \etal{} \cite{Vi19} and focus on five of the 14 textural features, namely homogeneity (H), contrast (CON), correlation (COR), local homogeneity (LH), and entropy (E):

\begin{align}
\textrm{H} &= \sum_{i=1}^{N}  \sum_{j=1}^{N} p_{\theta}(i,j)^2 \\
\textrm{CON} &= \sum_{i=1}^{N} \sum_{j=1}^{N} (i - j)^2 p_{\theta}(i,j) \\
\textrm{COR} &= \sum_{i=1}^{N}\sum_{j=1}^{N} \left( \frac{i - \mu_{x}}{\sigma_{x}} \right) \left(\frac{j - \mu_{y}}{\sigma_{y}} \right)p_{\theta}(i,j) \\ 
\textrm{LH} &= \sum_{i=1}^{N}  \sum_{j=1}^{N} \frac{1}{1+(i - j)^2}p_{\theta}(i,j) \\
\textrm{E} &= - \sum_{i=1}^{N}  \sum_{j=1}^{N}p_{\theta}(i,j)\ln(p_{\theta}(i,j)) \label{eq:OGE}
\end{align}
where $p_{x}(i)~=~\sum_{j=1}^{N}~p_{\theta}(i,j)$, $p_{y}(j)~=~\sum_{i=1}^{N}~p_{\theta}(i,j)$, $\mu_x~=~\sum_{i=1}^{N}i\cdot~p_{x}(i)$, $\mu_{y}~=~\sum_{j=1}^{N}j\cdot~p_{y}(j)$, $\sigma_{x}^2~=~\sum_{i=1}^{N}(i-\mu_x)^2\cdot~p_{x}(i)$, $\sigma_{y}^2~=~\sum_{j=1}^{N}(j-\mu_{y})^2\cdot~p_{y}(j)$. 

 A concise interpretation of the original Haralick measures is provided elsewhere in the literature \cite{Kr14,Vi19,MT22}. The \texttt{mahotas} package typically uses $\log_2$ for calculations of entropy, but has been modified to use the natural logarithm ($\ln$) in order to be consistent with the modified approach used within this work. This change results in a simple translation of the calculations of entropy, which is examined further in the supplementary material (section Entropy). 

\subsubsection*{Modified Haralick measures}\label{sec:methods:InvHaralick}
  
The source code, pulled from the \texttt{tomlof}  \href{https://github.com/tomlof/invariant-haralick-features}{github} \texttt{invariant-haralick-features} with commit \texttt{f83acbf}, is used to calculate both modified GLCM's ($\boldsymbol{\widetilde{p}}_{\theta}$, by dividing each element in $\textbf{P}_{\theta}$ by the sum, $\sum_{i,j=1}^{N}P_{\theta}(i,j)N^{-2}$) and all 14 modified textural measures. The amount of time required to compute textural measures is dependent on the magnitude of $N_{b}$ considered.  As an example, calculations using an $N_{b}$ of 50, 1000, and 10000 took an average of 0.21 $\pm$ 0.07, 2.0 $\pm$ 1.6, and 240 $\pm$ 73 seconds to complete on a single Intel Xeon processor (E5-2683). Calculated modified textural measures are averaged over all adjacency directions to compute directionally invariant measures. The work presented herein focuses on five modified textural features, namely homogeneity ($\widetilde{\mathrm{H}}$), contrast ($\widetilde{\mathrm{CON}}$), correlation ($\widetilde{\mathrm{COR}}$), local homogeneity ($\widetilde{\mathrm{LH}}$), and entropy ($\widetilde{\mathrm{E}}$): 

\begin{align}
\widetilde{\mathrm{H}} &= \sum_{i=1}^{N} \sum_{j=1}^{N} \widetilde{p}_{\theta}(i,j)^2 \Delta_{ij} \\
\widetilde{\mathrm{CON}} &= \sum_{i=1}^{N}\sum_{j=1}^{N} \left(\frac{i}{N} - \frac{j}{N}\right)^2 \widetilde{p}_{\theta}(i,j) \Delta_{ij} \\
\widetilde{\mathrm{COR}} &= \sum_{i=1}^{N}\sum_{j=1}^{N} \left( \frac{\frac{i}{N} - \widetilde{\mu}_{x}}{\widetilde{\sigma}_{x}} \right) \left( \frac{\frac{j}{N} - \widetilde{\mu}_{y}}{\widetilde{\sigma}_{y}} \right) \widetilde{p}_{\theta}(i,j) \Delta_{ij} \\ 
\widetilde{\mathrm{LH}} &= \sum_{i=1}^{N} \sum_{j=1}^{N} \frac{1}{1+\left(\frac{i}{N} - \frac{j}{N}\right)^2}\widetilde{p}_{\theta}(i,j)\Delta_{ij} \\
\widetilde{\mathrm{E}} &= - \sum_{i=1}^{N} \sum_{j=1}^{N}\widetilde{p}_{\theta}(i,j)\ln(\widetilde{p}_{\theta}(i,j))\Delta_{ij} \label{eq:modE}
\end{align}

\noindent
where $\Delta~=~\frac{1}{N}$, $\Delta_{ij}~=~\frac{1}{N^2}$, $\widetilde{p}_{x}(i)~=~\sum_{j=1}^{N}\widetilde{p}_{\theta}(i,j) \Delta$, $\widetilde{p}_{y}(j)~=~\sum_{i=1}^{N}\widetilde{p}_{\theta}(i,j) \Delta$, $\widetilde{\mu} _x~=~\sum_{i=1}^{N}\frac{i}{N}\cdot\widetilde{p}_{x}(i)$, $\widetilde{\mu} _{y}~=~\sum_{j=1}^{N}\frac{j}{N}\cdot \widetilde{p}_{y}(j)$, $\widetilde{\sigma}_{x}^2~=~\sum_{i=1}^{N} (\frac{i}{N}-\widetilde{\mu} _x)^2\cdot\widetilde{p}_{x}(i)$, $\widetilde{\sigma}_{y}^2~=~\sum_{j=1}^{N}(\frac{j}{N}-\widetilde{\mu} _{y})^2\cdot \widetilde{p}_{y}(j)$.

The modified Haralick features are renormalized/rescaled versions of the original features, and as noted by \Lof{} \etal{} \cite{Tl19} have interpretations which do not generally change when compared to the original measures. Of the features considered, the modified calculation of entropy is the only documented exception; the modified entropy measure can be negative \cite{Tl19}.

\subsection*{Analysis}\label{sec:meth:analysis}

Original and modified Haralick measures are computed for each 3D dosimetric dataset determined under TG186 and TG43 conditions for all 137 patients. The presentation of results focuses on modified Haralick measures in sections \nameref{sec:res:rawFeatures} and \nameref{sec:res:patientCohort}, while original and modified measures are compared in section \nameref{sec:res:compareApproaches}.  Alongside Haralick measures, some dose metrics are presented including $D90$ (minimum dose received by the hottest 90\% of the target volume), $D99$ (minimum dose received by the hottest 99\% of the target volume), and the Dose Homogeneity Index (DHI). Although different definitions of DHI exist in the literature \cite{ICRU58, Ti11}, we consider the following: 

\begin{equation}
	\mathrm{DHI} = \frac{V_{100} - V_{150}}{V_{100}} \label{eq:DHI}
\end{equation}
where $V_{100}$ and $V_{150}$ are the volumes of the prostate receiving  at least 100\% or 150\% of the nominal prescription dose (145 Gy), respectively.  

To investigate variations in results extracted from dose distributions determined under TG43 and TG186 conditions, we evaluate percent differences (\%$\Delta$) between extracted Haralick measures (generically denoted $f$)
\begin{equation}
\%\Delta_{f} = \frac{f_{\mathrm{TG43}} - f_{\mathrm{TG186}}}{f_{\mathrm{TG43}}} \cdot 100\% \label{eq:deltaFeature}
\end{equation}
and between D90 metrics
\begin{equation}
\%\Delta_{D90} = \frac{D90_{\mathrm{TG43}} - D90_{\mathrm{TG186}}}{D90_{\mathrm{TG43}}} \cdot 100\%. \label{eq:deltaD90}
\end{equation}

Trends in these metrics are examined using lines of best fit, calculated using the \texttt{curve\_fit()} function available within the \texttt{scipy} \cite{scipy} package for the Python programming language \cite{python}.

\section*{Results}\label{sec:results}

\subsection*{Haralick features for patient subset}\label{sec:res:rawFeatures}

\begin{figure}[H]
	\centering
\includegraphics[width=0.45\linewidth]{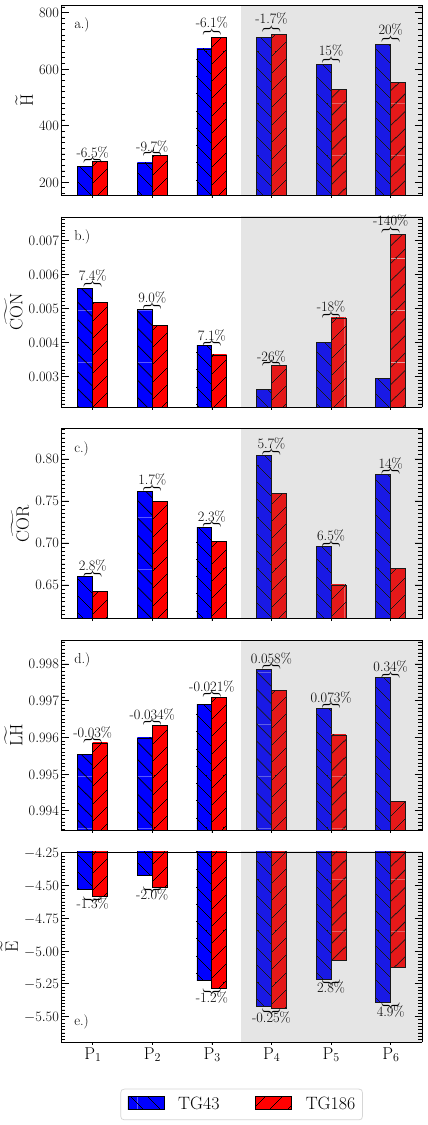}
	\caption{Modified Haralick measures for six patients (three with non-zero \%IC indicated by the shaded region of each panel). The percent difference \DDTG{} between textural measures extracted from TG43 and TG186 dose distributions is indicated adjacent to the respective bars within the figure. }
	\label{fig:subsetPatients}
\end{figure}

This section focuses on textural measures calculated using the modified Haralick approach for a 6 patient subset reflective of the larger patient cohort considered. Table~\ref{tab:subsetPatients}\mtab{tab:subsetPatients} provides \%IC as well as $D90,\, D99,$ and DHI for the 6 patients considered in this section. For the 6 patients, three have some level of intrapostatic calcification ranging between 0.56 and 3.8 \%, while the other three do not. The textural measures (figure~\ref{fig:subsetPatients})\mfig{fig:subsetPatients} vary with patient, reflecting variability in the 3D spatial distribution of absorbed doses due to \eg{} patient morphology, seed placement, and calculation conditions. Examples of absorbed dose distributions are shown in fig.~\ref{fig:GroupedJointdDVH_DoseDist}\mfig{fig:GroupedJointdDVH_DoseDist} (with further examples in Supplementary Materials figures 1 - 9).  Some trends observed in textural measures align with expectations based on clinical metrics \eg{} measures of homogeneity are largest for P$_4$ and smallest for P$_2$, and DHI is also largest for P$_4$ and smallest for P$_2$ out of the patient subset (table.~\ref{tab:subsetPatients}).

\begin{singlespace*}
\vspace{2em}
\begin{table}[H]
  \centering
  \caption{Summary of intraprostatic calcification (\%IC) and dosimetric data (TG43, TG186) for the subset of 6 patients considered in Results section \nameref{sec:res:rawFeatures}. }
    \begin{tabular}{p{0.1\textwidth}p{0.1\textwidth}p{0.1\textwidth}p{0.1\textwidth}p{0.1\textwidth}p{0.1\textwidth}p{0.1\textwidth}p{0.1\textwidth}} \toprule
          &       & \multicolumn{3}{c}{TG43} & \multicolumn{3}{c}{TG186} \\ \cmidrule(lr){3-5}\cmidrule(lr){6-8}
    patient & \multicolumn{1}{l}{\%IC} & \multicolumn{1}{l}{$D90$ (Gy)} & \multicolumn{1}{l}{$D99$ (Gy)} & \multicolumn{1}{l}{DHI} & \multicolumn{1}{l}{$D90$ (Gy)} & \multicolumn{1}{l}{$D99$ (Gy)} & \multicolumn{1}{l}{DHI} \\ \midrule
    P$_{1}$ & 0     & 150   & 80  & 0.22 & 140   & 75 & 0.25\\
    P$_{2}$ & 0     & 120   & 67  & 0.17 & 100   & 60 & 0.18\\
    P$_{3}$ & 0     & 140   & 88  & 0.40 & 130   & 83 & 0.45\\
    P$_{4}$ & 0.56  & 93    & 60  & 0.54 & 85    & 54 & 0.57\\
    P$_{5}$ & 1.6   & 150   & 104 & 0.42 & 140   & 96 & 0.45\\
    P$_{6}$ & 3.8   & 140   & 98  & 0.36 & 120   & 85 & 0.50\\ \bottomrule
    \end{tabular}%
  \label{tab:subsetPatients}%
\end{table}%
\end{singlespace*}


%
%

\begin{figure}[H]
	\centering
\includegraphics[width=\linewidth]{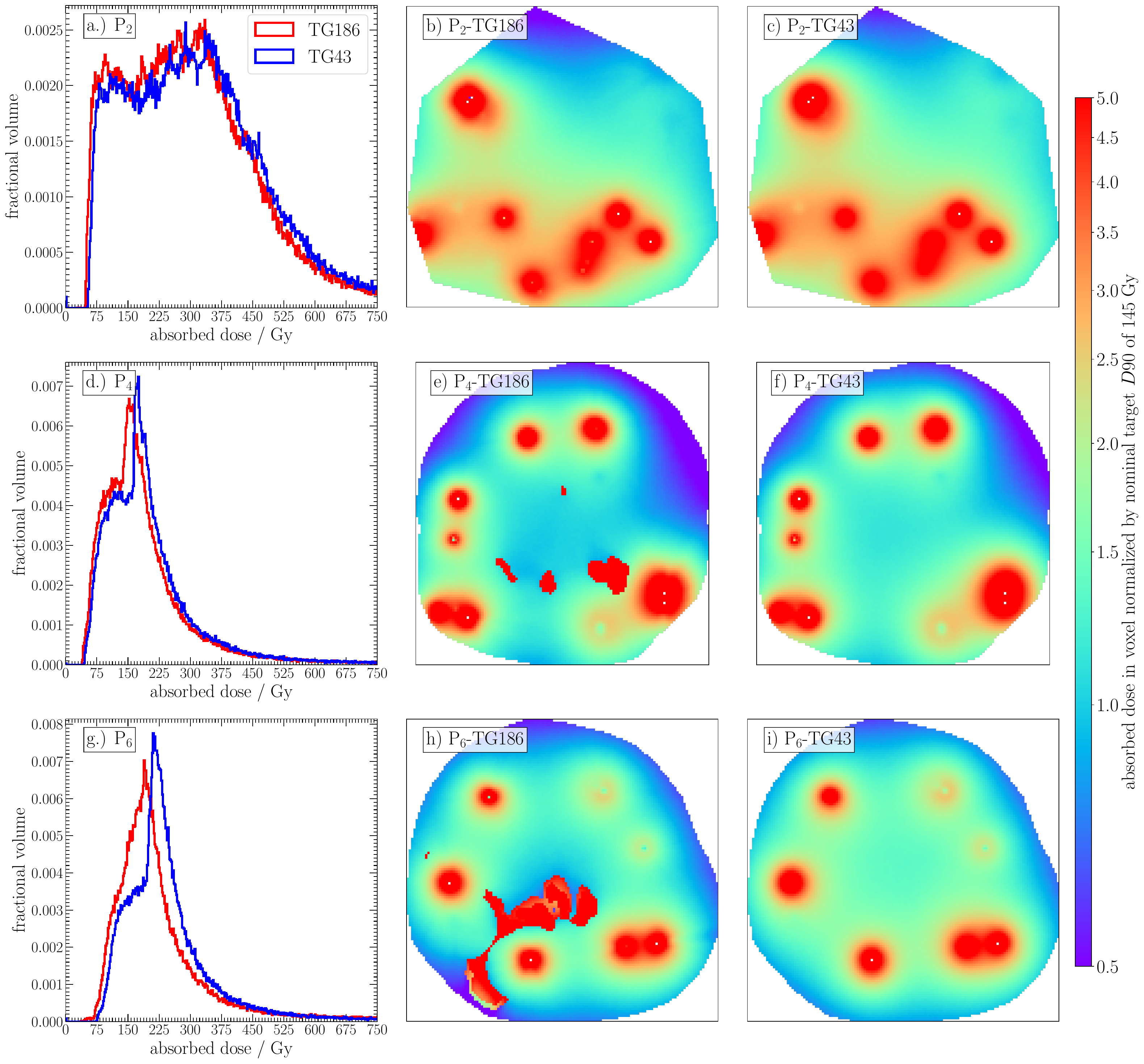}
	\caption[Differential Dose Volume Histograms and cross section of central 2D absorbed dose distributions calculated under TG43 and TG186 conditions]{Dose distributions for example patients P$_{2}$, P$_{4}$, and P$_{6}$: (a,d,g) differential Dose Volume Histograms;  central plane 2D absorbed doses calculated under (b,e,h) TG186 and (c,f,i) TG43 conditions (with doses relative to the 145~Gy prescription dose indicated by color). Intraprostatic calcifications are visible in (e) and (h) as contiguous `red' regions corresponding to doses much larger than the prescription dose.}
	\label{fig:GroupedJointdDVH_DoseDist}
\end{figure}

Considering patient P$_2$, measures of homogeneity are relatively low while measures of entropy are highest (small and negative) for either TG43 or TG186 conditions relative to others in the subset (figure \ref{fig:subsetPatients}a,e). These trends in textural measures reflect relatively sporadic absorbed dose distributions and a wide range of adjacency pair combinations (\ie{} distinct combinations of dose deposited across adjacent scoring voxels) caused by the patient morphology and relative seed placement (but not intraprostatic calcification as \%IC$=0$). These observations are supported by calculated magnitudes of $D90$, $D99$, DHI (tab.~\ref{tab:subsetPatients}) and dDVHs. In particular, Fig.~\ref{fig:GroupedJointdDVH_DoseDist}a shows relatively wide dDVHs for P$_2$ which also correspond to large values of the ratio  $D90/D99$ relative to others in the patient cohort: doses to voxels in the target vary considerably for this example patient, quantitatively characterized by relatively low homogeneity and high entropy measures compared with others in the patient subset.

Relatively low measures of entropy and contrast in conjunction with relatively high measures of homogeneity, correlation, and local homogeneity are observed for P$_4$ relative to others in the subset. These trends in textural measures reflect the (TG43, TG186) dose distributions for this patient which are relatively uniform with a narrow range of adjacency pair combinations and generally small differences in magnitudes of dose in adjacent voxels. Qualitative differences in the spatial distribution of absorbed dose deposited may be examined using cross sectional arrays of 3D absorbed dose distributions presented in figure~\ref{fig:GroupedJointdDVH_DoseDist} (panels b,e,h and c,f,i for TG186 and TG43 conditions, respectively). Examination of figure~\ref{fig:GroupedJointdDVH_DoseDist}e,f  shows that there are large cold spots at the periphery of the target volume for P$_4$. These observations are supported by relatively low magnitudes of $D90$ and $D99$, but high magnitudes of DHI (all relative to others in the subset).

The Haralick measures quantitatively characterize differences in dose distributions computed under TG43 and TG186 conditions. Comparing dDVHs (fig.~\ref{fig:GroupedJointdDVH_DoseDist}a,d,g) calculated using TG43 and TG186 conditions for the patient subset reveals dosimetric effects induced by differences in material composition and interseed attenuation. For patients without intraprostatic calcification (P$_{1}$, P$_2$, P$_3$),  doses computed under TG186 conditions have a lower overall doses with the dDVH shifted to lower doses compared to TG43 (fig.~\ref{fig:GroupedJointdDVH_DoseDist}a). For patient with intraprostatic calcification  (P$_{4}$,  P$_5$,  P$_6$), doses are further decreased in comparing TG186 relative to TG43 (fig.~\ref{fig:GroupedJointdDVH_DoseDist}d,g). Textural measures quantify these observed differences in the spatial distribution of absorbed dose. For example, measures of contrast increase with increasing \%IC for TG186 calculations due to large dosimetric differences in voxels containing prostate tissue compared with calcification, quantifying the abrupt changes in patterns of absorbed dose.

The values of the percent difference $\%\Delta_f$ between textural measures extracted from TG43 and TG186 dose distributions (eq.~\ref{eq:deltaFeature}) are shown for each Haralick measure in figure~\ref{fig:subsetPatients}. Magnitudes of \DDTG{} increase as patient intraprostatic calcification (\%IC) increases, thus quantifying the considerable effects of calcifications on the spatial distribution of TG186 doses, but not for TG43 doses (as calcifications are not considered).  For patients without intraprostatic calcification, \DDTG{} is relatively small ($<$10\% for the subset of patients presented). Correspondingly, TG43 and TG186 dose distributions for patients with 0 \%IC (P$_{1}$, P$_2$, P$_3$) differ less dramatically, with material composition and interseed attenuation resulting in dose reductions in regions adjacent to sources for TG186 compared to TG43 (see, \eg, Supplementary materials fig.~9 presents voxel-by-voxel comparison of cross sectional arrays from 3D absorbed dose distributions calculated using TG43 and TG186 conditions; panels a,b,c present patients with 0 \%IC, d,e,f present patients with non-zero \%IC).  Patient P$_{6}$ has the largest \%IC of the patient subset and, correspondingly, the largest magnitudes of \DDTG{} for each texture measure.

\subsection*{Haralick analysis of patient cohort}\label{sec:res:patientCohort}

\begin{figure}[H]
	\centering
\includegraphics[height =0.9\textheight]{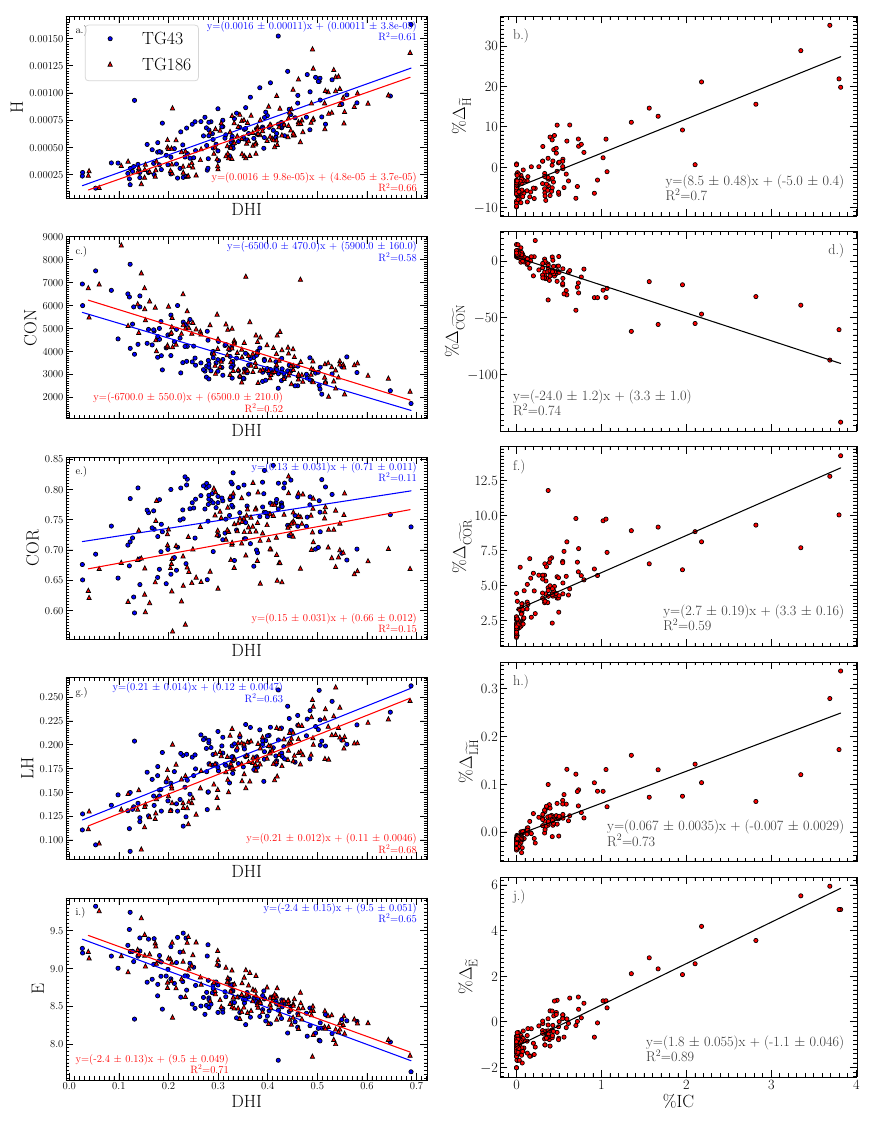}
	\caption[Textural measures extracted from patient specific PIPB absorbed dose distributions as a function of Dose Homogeneity Index (DHI) and percent difference between calculations using TG43 and TG186 conditions as a function of intraprostatic calcification percent by total prostate volume (\%IC)]{Textural measures extracted from patient specific PIPB absorbed dose distributions as a function of (a,c,e,g,i) Dose Homogeneity Index (DHI; eq.\ref{eq:DHI}) and (b,d,f,h,j) percent difference between calculations using TG43 and TG186 conditions as a function of intraprostatic calcification percent by total prostate volume (\%IC)}
	\label{fig:ICdiff_DHI}
\end{figure}

Figure~\ref{fig:ICdiff_DHI}\mfig{fig:ICdiff_DHI} presents modified Haralick measures extracted from dose distributions for the full \NPatients{} patient cohort as a function of DHI and \%IC; in complement, additional figures for original and modified textural measures are provided in Supplementary Materials.

There are considerable variations in textural measures as well as DHI across the full patient cohort, reflecting the distinct spatial distributions of absorbed dose for each patient.  The textural measures quantify the 3D dose distributions in voxels, in contrast with DHI which quantifies the target volumes receiving 100 and 150\% of the prescription dose without consideration of the detailed spatial pattern of absorbed dose in voxels.  Some of the textural measures are reasonably well correlated with DHI, although there is spread about the lines of best fit shown in fig.~\ref{fig:ICdiff_DHI}; coefficients of determination ($R^2$) range between 0.52 and 0.71 for homogeneity, contrast, local homogeneity, and entropy (fig.~\ref{fig:ICdiff_DHI} a,c,g,i) but is smaller for the correlation measure (fig.~\ref{fig:ICdiff_DHI} e). Increasing values of DHI correspond to the differences between $V_{150}$ and $V_{100}$ becoming smaller (by definition -- eq.~\ref{eq:DHI}) and the dose distribution becoming more uniform. Correspondingly, there will be smaller differences in the magnitude of dose deposited across adjacent  voxels, and recurring patterns of absorbed dose become more common, resulting in measures of homogeneity and local homogeneity increasing as DHI increases (fig.~\ref{fig:ICdiff_DHI}a,g), but decreasing contrast and entropy (fig.~\ref{fig:ICdiff_DHI}e,i). Other first order measures considered of $D90$ and $D99$ are observed to vary independently from calculated textural measures (Supplementary Materials section Textural measures extracted from patient specific 3D absorbed dose distributions; fig. 11, 12).

For all Haralick measures, there is an overall trend of \DDTG{} increasing in magnitude with increasing \%IC (fig.~\ref{fig:ICdiff_DHI} b, d, f, h and j), consistent with observations for the patient subset (Results section \nameref{sec:res:rawFeatures}). A strong correlation between \DDTG{} and \%IC ($R^2$ between 0.59 and 0.9) is observed for all texture measures, with some scatter about the line of best fit.  In comparison to the trends between \DDTG{} and \%IC, large variations in each textural measure are observed between patients, consistent with Results section \nameref{sec:res:rawFeatures} and shown in Supplementary Materials fig.~13 and 14 
Thus, \DDTG{} facilitates a focused analysis for the comparison of calculations performed under TG43 and TG186 conditions. 

For patients with 0 \%IC, differences between dose distributions calculated under TG43 and TG186 are relatively small, thus values of \DDTG{} are relatively small (fig.~\ref{fig:ICdiff_DHI} b,d,f,h,j; fitted y-intercept: $<$ 5\%). As noted above, the magnitude of \DDTG{} increases with \%IC: \DDTG{} increases with \%IC for homogeneity, correlation, local homogeneity, and entropy;  \DDTG{} for contrast decreases as a function of increasing \%IC. These trends correspond to TG186 dose distributions become more sporadic, with large differences in dose magnitudes in adjacent voxels and generally larger variations in the adjacency pair combinations as \%IC increases (relative to TG43 doses).

The Supplementary Materials figure~14 presents \DDTG{} between Haralick measures calculated under TG43 and TG186 conditions as a function of \DDninty{} (the percent difference in calculations of $D90$ for absorbed dose distributions calculated under TG43 and TG186 conditions; eq.~\ref{eq:deltaD90}). Similar trends between \DDTG{} and \DDninty{} to the trends reported between \DDTG{} and \%IC are observed. Generally, a weaker correlation is measured by coefficients of determination ($R^2$) between extracted textural measures and \DDninty{} (0.31 - 0.44) when compared to \%IC (0.59 - 0.90). 

\subsection*{Comparison of original and modified textural analysis}\label{sec:res:compareApproaches}

\begin{figure}[H]
	\centering
\includegraphics[width=0.45\linewidth]{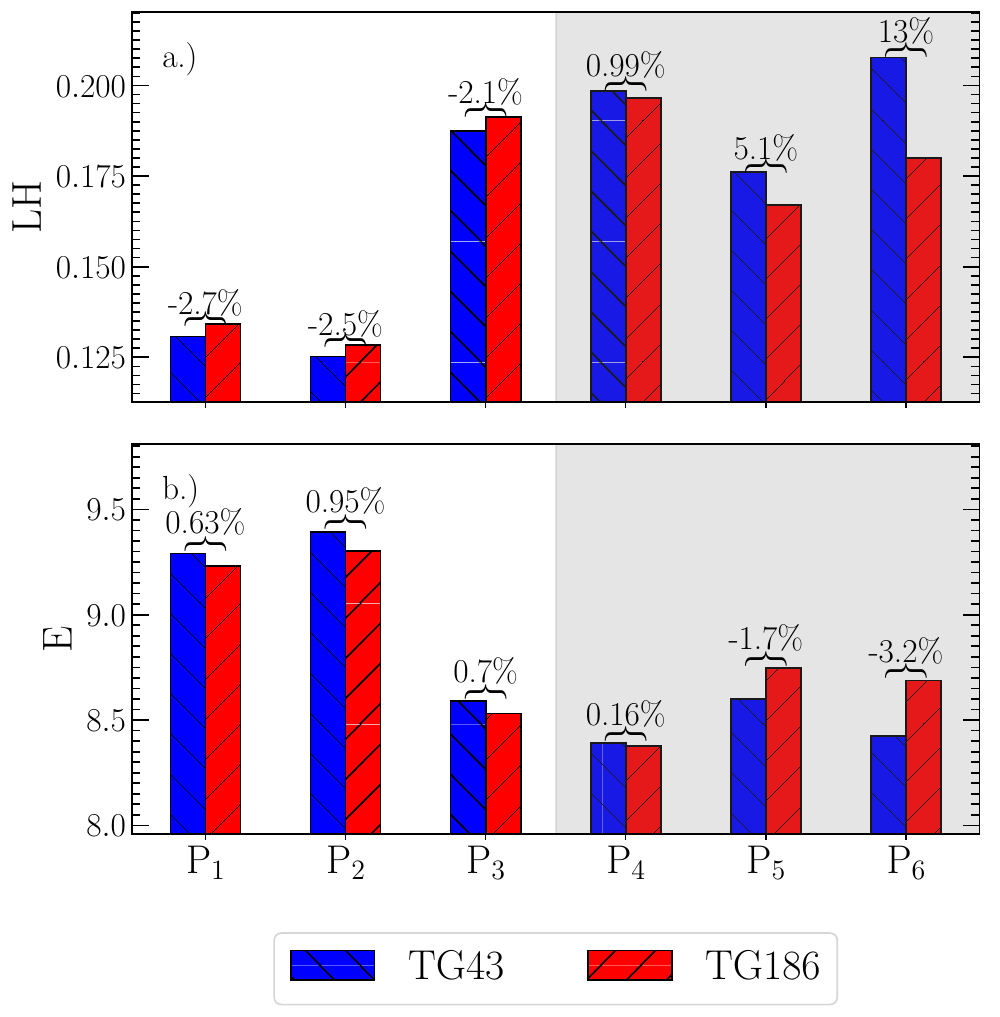}
	\caption{Original Haralick measures for six patients (three with non-zero \%IC indicated by the shaded region of each panel; tab.~\ref{tab:subsetPatients}). The percent difference \DDTG{} between textural measures extracted from TG43 and TG186 dose distributions is indicated adjacent to the respective bars within the figure.}
	\label{fig:subsetPatientsCompare}
\end{figure}

Figure~\ref{fig:subsetPatientsCompare}\mfig{fig:subsetPatientsCompare} presents original measures of local homogeneity and entropy for the six-patient subset, for comparison with the modified Haralick measures presented in Results section~\nameref{sec:res:rawFeatures}. Measures of homogeneity and contrast are omitted because they are related via a simple scaling; measures of correlation are omitted because it is mathematically identical in the original and modified approaches  (see Supplementary Materials section Comparison of original and modified textural measures for mathematical relationships between original and modified Haralick measures).

When comparing original and modified textural measures for the subset of patient data presented in figures~\ref{fig:subsetPatients} and \ref{fig:subsetPatientsCompare}, differences in both the magnitude and trends of extracted measures are observed. Notable differences in both the magnitudes of extracted features and trends across patients are observed between original and modified measures of local homogeneity and entropy. Differences between calculations of entropy appear as a linear shift, while differences between original and modified calculations of local homogeneity do not have a clear relationship. 

\begin{figure}[H]
	\centering
	\includegraphics[height =0.9\textheight]{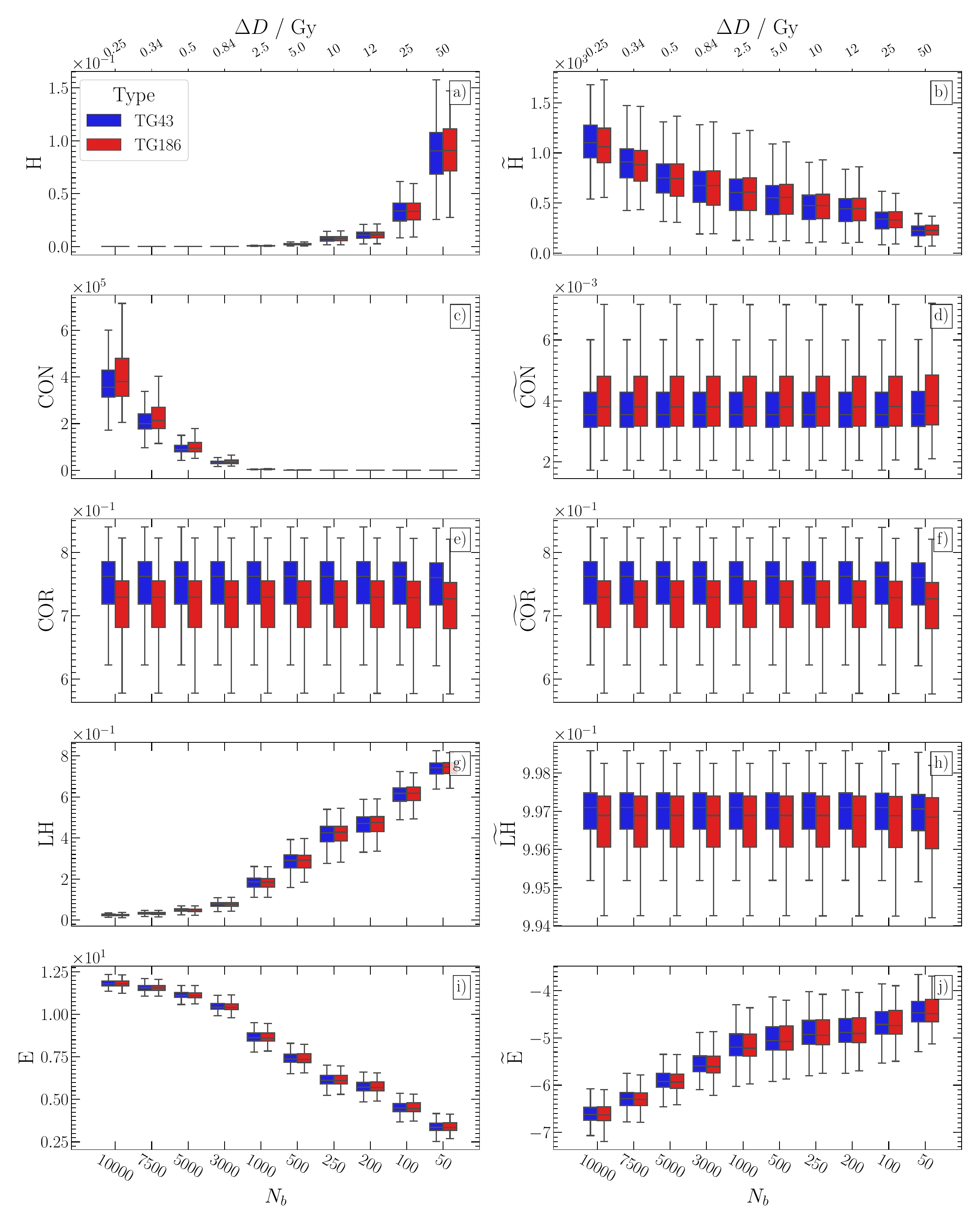}
	\caption[Original and modified textural features for all \NPatients{} patients calculated under aggregated based on calculation conditions (TG43 and TG186) as a function of quantization level width ($\Delta D$).]{Sensitivity of textural measures to quantization level width ($\Delta D$) for (a,c,e,g,i) Original and (b,d,f,h,j) modified textural features considering all \NPatients{} patients calculated under aggregated based on calculation conditions (TG43 and TG186). The box extends between the upper and lower quartiles; its horizontal line indicates the median.}
	\label{fig:Sensitivity}
\end{figure}

While previous sections considered a fixed $N_b=1000$ ($\Delta D=$ 2.5 Gy) for the Haralick analysis, in the following we investigate the sensitivity of original and modified Haralick measures to $N_b$ with corresponding quantization level widths $\Delta D$ between 0.25 and 50 Gy (tab.~\ref{tab:deltaD}). Figure~\ref{fig:Sensitivity}\mfig{fig:Sensitivity} presents original and modified Haralick measures collected from all \NPatients{} 3D patient specific absorbed dose distributions considering different $\Delta D$, grouped based on calculations performed under either TG43 or TG186 conditions (Supplementary materials section Sensitivity of relative trends to quantization bin width provides additional figures, presenting each feature as a function of \%IC calculated using different values of $\Delta D$; fig.15-19). Both original and modified Haralick measures are observed to be sensitive to quantization level width. 
Larger differences in calculated original Haralick measures induced by changes to $\Delta D$ are observed when compared to modified Haralick measures. 
In particular, measures of homogeneity and local homogeneity increase while measures of contrast and entropy decrease as the quantization level width increases. Although modified Haralick measures are relatively less sensitive to changes in $\Delta D$, trends in calculated modified textural measures as a function of $\Delta D$ are observed. Measures of homogeneity decrease, and measures of entropy are increase as larger values of $\Delta D$ are considered.

\section*{Discussion}\label{sec:disc}

Haralick analysis provides a quantitative characterization of the 3D spatial distribution of doses from patient-specific MC simulation. Our results, for Haralick texture measures extracted from PIPB dose distributions computed under TG43 or TG186 conditions, show large intrafeature differences which reflect the variability in the spatial distribution of absorbed doses for these patient specific data. Textural measures are sensitive to dosimetric differences due to patient morphology, seed position and density, material/tissue compositions, and interseed attenuation.

The presented work demonstrates that Haralick texture features extracted from patient specific dosimetric data can be interpreted in terms of fundamental dosimetry. Our work suggests that supplementing existing analysis techniques reliant on first order measures (\eg{} $D90$, DHI) with textural analysis provides an enhanced approach to characterize 3D patient specific absorbed dose distributions. The work presented is distinct from recent applications of texture analysis in the emergent field of dosiomics which involves correlating extracted measures (textural and traditional Dose Volume Histogram (DVH) derived measures) to radiotherapy outcomes \cite{Gh18,Em21,Pl21}. Where dosiomic studies generally omit the presentation and interpretation of calculated measures, this is an explicit objective of the study presented.



Intraprostatic calcifications are observed to be directly associated with measurable textural changes within absorbed dose distributions. Examination of the relative measure \DDTG{} (the percent difference between textural measures extracted from absorbed dose distributions generated under TG43 and TG186 conditions, eq.~\ref{eq:deltaFeature}) facilitates focused analysis on the changes introduced with varying levels of \%IC. For patients with 0 \%IC, relatively small intrafeature differences between textural measures for TG43 and TG186 doses are observed, corresponding to the relatively small dosimetric differences associated primarily with the effects of interseed attenuation modelling and soft tissue composition \cite{Mi17, Zk15} (fig.~\ref{fig:GroupedJointdDVH_DoseDist} b,c). As dose distributions for patients with larger \%IC are examined, linear trends between \DDTG{} for all textural measures and \%IC are observed (coefficients of determination $R^2$ range between 0.59 and 0.9 which are comparable in magnitude to those calculated by Miksys \etal{} \cite{Mi17} when investigating the relationship between $D90$ and \%IC.). The trends are reflecting increasingly large differences in the spatial patterns of absorbed dose within 3D absorbed dose distributions calculated under TG186 conditions relative to TG43 conditions as patients with higher \%IC are evaluated. 

Most results are presented for the quantization bin width $\Delta D$ = 2.5 Gy ($N_b$ = 1000) which corresponds to $<2\%$ of the nominal prescription dose of 145~Gy. As shown in Results (section ~\nameref{sec:res:compareApproaches}), although original measures are relatively more sensitive, both original and modified measures are sensitive to the quantization level bin width. The linear quantization scheme (eq.~\ref{eq:mapping}) results in all quantized distributions of absorbed dose ($\mathcal{D'}$) having identical values of $N$ ($N = N_b$). Similar trends are typically observed when comparing both original and modified measures (\eg{} for the subset of example patients: fig.~\ref{fig:subsetPatients} and \ref{fig:subsetPatientsCompare}), and differences in either the magnitude or trends in extracted features  are the result of the rescaling/renormalizations used for the modified textural measures. For the work presented herein, both original and modified textural measures are demonstrated to be capable of quantifying dosimetric differences between patients. For example, with consideration of expected changes induced by either original and modified measures (\eg{} linear shift of entropy measure) comparison of \DDTG{} calculated using original and modified measures as a function of \%IC reveals comparable trends and measures of $R^2$ (Supplementary Materials fig.~13). Future work applying Haralick analysis to dosimetric data may necessitate alternative approaches to quantization and may demonstrate stronger benefits for the use of the modified approach to texture analysis



Haralick texture analysis is demonstrated to enhance existing analysis methodologies used to examine PIPB dosimetric data. Traditional dose metrics such as $D90$ are incapable of quantifying heterogeneities which may exist within absorbed dose distributions \cite{Na00}. Therefore Haralick analysis may provide large clinical value in future work processing, distilling, and sharing patient specific data. Many, interesting possibilities for future work remain to examine patient specific distributions, such as building on the work presented herein to characterize different brachytherapy sources based on activity, design, and placement. Furthermore, future work will necessitate examination of clinical integration for Haralick textural measures into treatment planning and evaluation. Such integrations may provide the possibility of yielding new insights from patient specific absorbed dose distributions by examining organ at risk volumes in addition to target volumes in conjunction with relating to treatment outcome data. 

\section*{Conclusion}

The present work demonstrates that Haralick texture analysis can quantitatively characterize 3D patient-specific absorbed dose distributions of low dose rate perminant implant prostate brachytherapy determined via MC simulation. Original and modified Haralick measures of homogeneity, contrast, correlation, local homogeneity, and entropy characterize differences in 3D absorbed dose distributions due to simulation conditions (TG43 versus TG186) as well as the effects of intraprostatic calcification. The quantization approach considered for the work presented herein generally results in linear transformations (scaling and translations) of modified relative to original Haralick measures. The trends in Haralick measures considered, namely homogeneity, contrast, correlation, local homogeneity, and entropy, are interpretable in terms of the underlying radiation physics. Interesting possibilities remain for future work, including relating to biologically-motivated quantities and ensuing biological response.


\section*{Acknowledgements}
The authors acknowledge support from the Natural Sciences and Engineering Research Council of Canada (NSERC) [funding reference number 06267-2016], NSERC Postgraduate Scholarship Doctoral (PGS-D) [funding reference number 546549-2020], Canada Research Chairs (CRC) program, Compute/Calcul Canada, and the Kiwanis Club of Ottawa Medical Foundation and Dr. Kanta Marwah Scholarship in Medical Physics.

\clearpage

\section*{References}
\addcontentsline{toc}{section}{\numberline{}References}






\begin{thebibliography}{}

\end{thebibliography}


\vspace*{-20mm}

\begin{thebibliography}{10}

\bibitem{ICRU85}
ICRU,
\newblock {Report 85: Fundamental Quantities and Units for Ionizing Radiation},
\newblock J. of the ICRU {\bf 11} (2011).

\bibitem{TG105}
I.~J. Chetty, B.~Curran, J.~E. Cygler, J.~J. DeMarco, G.~Ezzell, B.~A.
  Faddegon, I.~Kawrakow, P.~J. Keall, H.~Liu, C.-M.~C. Ma, D.~W.~O. Rogers,
  J.~Seuntjens, D.~Sheikh-Bagheri, and J.~V. Siebers,
\newblock {Report of the AAPM Task Group No. 105: Issues associated with
  clinical implementation of Monte Carlo-based photon and electron external
  beam treatment planning},
\newblock Med. Phys. {\bf 34}, 4818--4853 (2007).

\bibitem{TG106}
I.~J. Das, C.-W. Cheng, R.~J. Watts, A.~Ahnesjö, J.~Gibbons, X.~A. Li,
  J.~Lowenstein, R.~K. Mitra, W.~E. Simon, and T.~C. Zhu,
\newblock {Accelerator beam data commissioning equipment and procedures: Report
  of the TG-106 of the Therapy Physics Committee of the AAPM},
\newblock Med. Phys. {\bf 35}, 4186--4215 (2008).

\bibitem{TG185}
J.~B. Farr, M.~F. Moyers, C.~E. Allgower, M.~Bues, W.-C. Hsi, H.~Jin, D.~N.
  Mihailidis, H.-M. Lu, W.~D. Newhauser, N.~Sahoo, R.~Slopsema, D.~Yeung, and
  X.~R. Zhu,
\newblock {Clinical commissioning of intensity-modulated proton therapy
  systems: Report of AAPM Task Group 185},
\newblock Med. Phys. {\bf 48}, e1--e30 (2021).

\bibitem{TG186}
L.~Beaulieu, {\r A}.~Carlsson~Tedgren, J.-F. Carrier, S.~D. Davis, F.~Mourtada,
  M.~J. Rivard, R.~M. Thomson, F.~Verhaegen, T.~A. Wareing, and J.~F.
  Williamson,
\newblock {Report of the Task Group 186 on model-based dose calculation methods
  in brachytherapy beyond the TG-43 formalism: Current status and
  recommendations for clinical implementation},
\newblock Med. Phys. {\bf 39}, 6208--6236 (2012).

\bibitem{Ev19}
E.~Vigneault, K.~Mbodji, D.~Carignan, A.-G. Martin, N.~Miksys, R.~M. Thomson,
  S.~Aubin, N.~Varfalvy, and L.~Beaulieu,
\newblock {The association of intraprostatic calcifications and dosimetry
  parameters with biochemical control after permanent prostate implant},
\newblock Brachytherapy {\bf 18}, 787--792 (2019).

\bibitem{Mi17}
N.~Miksys, E.~Vigneault, A.-G. Martin, L.~Beaulieu, and R.~M. Thomson,
\newblock {Large-scale retrospective Monte Carlo dosimetric study for permanent
  implant prostate brachytherapy},
\newblock Int. J. Radiat. Oncol. Biol. Phys. {\bf 97}, 606--615 (2017).

\bibitem{Ma04}
M.~H. Bharati, J.~Liu, and J.~F. MacGregor,
\newblock {Image texture analysis: methods and comparisons},
\newblock Chemometrics and Intelligent Laboratory Systems {\bf 72}, 57--71
  (2004).

\bibitem{MM98}
A.~Materka and M.~Strzelecki,
\newblock {Texture analysis methods - A review},
\newblock COST B11 Report , 1 – 33 (1998).

\bibitem{ST99}
L.-K. Soh and C.~Tsatsoulis,
\newblock {Texture analysis of {SAR} sea ice imagery using gray level
  co-occurrence matrices},
\newblock IEEE Transactions on Geoscience and Remote Sensing {\bf 37}, 780--795
  (1999).

\bibitem{Ca01}
D.~A. Clausi,
\newblock {Comparison and fusion of co‐occurrence, {Gabor} and {MRF} texture
  features for classification of {SAR} sea‐ice imagery},
\newblock Atmosphere-Ocean {\bf 39}, 183--194 (2001).

\bibitem{Dp12}
P.~Dollar, C.~Wojek, B.~Schiele, and P.~Perona,
\newblock {Pedestrian Detection: An Evaluation of the State of the Art},
\newblock IEEE Transactions on Pattern Analysis and Machine Intelligence {\bf
  34}, 743--761 (2012).

\bibitem{Zh19}
Y.~Zhang, T.~E. Plautz, Y.~Hao, C.~Kinchen, and X.~A. Li,
\newblock {Texture-based, automatic contour validation for online adaptive
  replanning: A feasibility study on abdominal organs},
\newblock Med. Phys. {\bf 46}, 4010--4020 (2019).

\bibitem{Vi19}
I.~Vrbik, S.~J. Van~Nest, P.~Meksiarun, J.~Loeppky, A.~Brolo, J.~J. Lum, and
  A.~Jirasek,
\newblock {Haralick texture feature analysis for quantifying radiation response
  heterogeneity in murine models observed using {Raman} spectroscopic mapping},
\newblock PLoS One {\bf 14}, 1--12 (2019).

\bibitem{Vm15}
M.~Valli{\`{e}}res, C.~R. Freeman, S.~R. Skamene, and I.~E. Naqa,
\newblock {A radiomics model from joint {FDG}-{PET} and {MRI} texture features
  for the prediction of lung metastases in soft-tissue sarcomas of the
  extremities},
\newblock Phys. Med. Biol. {\bf 60}, 5471--5496 (2015).

\bibitem{Rl18}
L.~Rossi, R.~Bijman, W.~Schillemans, S.~Aluwini, C.~Cavedon, M.~Witte,
  L.~Incrocci, and B.~Heijmen,
\newblock {Texture analysis of {3D} dose distributions for predictive modelling
  of toxicity rates in radiotherapy},
\newblock Radiotherapy and Oncology {\bf 129}, 548--553 (2018).

\bibitem{Li19}
B.~Liang, H.~Yan, Y.~Tian, X.~Chen, L.~Yan, T.~Zhang, Z.~Zhou, L.~Wang, and
  J.~Dai,
\newblock {Dosiomics: Extracting 3D Spatial Features From Dose Distribution to
  Predict Incidence of Radiation Pneumonitis},
\newblock Frontiers in Oncology {\bf 9} (2019).

\bibitem{Ta21}
T.~Adachi, M.~Nakamura, T.~Shintani, T.~Mitsuyoshi, R.~Kakino, T.~Ogata,
  T.~Ono, H.~Tanabe, M.~Kokubo, T.~Sakamoto, Y.~Matsuo, and T.~Mizowaki,
\newblock {Multi-institutional dose-segmented dosiomic analysis for predicting
  radiation pneumonitis after lung stereotactic body radiation therapy},
\newblock Med. Phys. {\bf 48}, 1781--1791 (2021).

\bibitem{MT22}
I.~R. Mansour and R.~M. Thomson,
\newblock {Haralick texture feature analysis for characterization of specific
  energy and absorbed dose distributions across cellular to patient length
  scales},
\newblock Phys. Med. Biol. {\bf 68}, 075006 (2023).

\bibitem{Ha73}
R.~M. Haralick, K.~Shanmugam, and I.~Dinstein,
\newblock {Textural Features for Image Classification},
\newblock IEEE Transactions on Systems, Man, and Cybernetics {\bf 3}, 610--621
  (1973).

\bibitem{Tl19}
T.~L{\"o}fstedt, P.~Brynolfsson, T.~Asklund, T.~Nyholm, and A.~Garpebring,
\newblock {Gray-level invariant Haralick texture features},
\newblock PLoS One {\bf 14}, e0212110 (2019).

\bibitem{Bp17}
P.~Brynolfsson, D.~Nilsson, T.~Torheim, T.~Asklund, C.~T. Karlsson, J.~Trygg,
  T.~Nyholm, and A.~Garpebring,
\newblock {Haralick texture features from apparent diffusion coefficient
  ({ADC}) {MRI} images depend on imaging and pre-processing parameters},
\newblock Scientific Reports {\bf 7}, 4041 (2017).

\bibitem{Ca02}
D.~A. Clausi,
\newblock {An analysis of co-occurrence texture statistics as a function of
  grey level quantization},
\newblock Canadian Journal of Remote Sensing {\bf 28}, 18 (2002).

\bibitem{Dg17}
H.~Di and D.~Gao,
\newblock {Nonlinear gray-level co-occurrence matrix texture analysis for
  improved seismic facies interpretation},
\newblock Interpretation {\bf 5}, SJ31--SJ40 (2017).

\bibitem{Rl15}
R.~T. Leijenaar, G.~Nalbantov, S.~Carvalho, W.~J. van Elmpt, E.~G. Troost,
  R.~Boellaard, H.~J. Aerts, R.~J. Gillies, and P.~Lambin,
\newblock {The effect of {SUV} discretization in quantitative {FDG}-{PET}
  {Radiomics}: the need for standardized methodology in tumor texture
  analysis},
\newblock Scientific Reports {\bf 5}, 11075 (2015).

\bibitem{Gw12}
W.~Gomez, W.~C.~A. Pereira, and A.~F.~C. Infantosi,
\newblock {Analysis of Co-Occurrence Texture Statistics as a Function of
  Gray-Level Quantization for Classifying Breast Ultrasound},
\newblock IEEE Transactions on Medical Imaging {\bf 31}, 1889--1899 (2012).

\bibitem{Va17}
M.~Vallières, E.~Kay-Rivest, L.~J. Perrin, X.~Liem, C.~Furstoss, H.~J. W.~L.
  Aerts, N.~Khaouam, P.~F. Nguyen-Tan, C.-S. Wang, K.~Sultanem, J.~Seuntjens,
  and I.~El~Naqa,
\newblock {Radiomics strategies for risk assessment of tumour failure in
  head-and-neck cancer},
\newblock Scientific Reports {\bf 7}, 10117 (2017).

\bibitem{Ga18}
A.~Garpebring, P.~Brynolfsson, P.~Kuess, D.~Georg, T.~H. Helbich, T.~Nyholm,
  and T.~Löfstedt,
\newblock {Density estimation of grey-level co-occurrence matrices for image
  texture analysis},
\newblock Phys. Med. Biol. {\bf 63}, 195017 (2018).

\bibitem{MT23}
I.~R. Mansour and R.~M. Thomson,
\newblock Haralick texture analysis for microdosimetry: characterization of
  Monte Carlo generated 3D specific energy distributions,
\newblock Physics in Medicine \& Biology {\bf 68}, 185003 (2023).

\bibitem{Na00}
S.~Nag, W.~Bice, K.~DeWyngaert, B.~Prestidge, R.~Stock, and Y.~Yu,
\newblock {The American Brachytherapy Society recommendations for permanent
  prostate brachytherapy postimplant dosimetric analysis},
\newblock Int. J. Radiat. Oncol. Biol. Phys. {\bf 46}, 221--230 (2000).

\bibitem{Ma14}
J.~Mason, B.~Al-Qaisieh, P.~Bownes, A.~Henry, and D.~Thwaites,
\newblock Investigation of interseed attenuation and tissue composition effects
  in 125I seed implant prostate brachytherapy,
\newblock Brachytherapy {\bf 13}, 603--610 (2014).

\bibitem{Sa13}
H.~Safigholi, D.~Sardari, S.~K. Jashni, S.~R. Mahdavi, and A.~S. Meigooni,
\newblock An analytical model to determine interseed attenuation effect in
  low-dose-rate brachytherapy,
\newblock J. of App. Clin. Med. Phys. {\bf 14}, 150--163 (2013).

\bibitem{egsnrc}
I.~Kawrakow, D.~Rogers, E.~Mainegra-Hing, F.~Tessier, R.~Townson, and
  B.~Walters,
\newblock {EGSnrc toolkit for Monte Carlo simulation of ionizing radiation
  transport},
\newblock \href{https://doi.org/10.4224/40001303}{doi:10.4224/40001303}
  [release v2021], 2021.

\bibitem{Ta07}
R.~E.~P. Taylor, G.~Yegin, and D.~W.~O. Rogers,
\newblock {Benchmarking BrachyDose: Voxel based EGSnrc Monte Carlo calculations
  of TG-43 dosimetry parameters},
\newblock Medical Physics {\bf 34}, 445--457 (2007).

\bibitem{Mi15}
N.~Miksys, C.~Xu, L.~Beaulieu, and R.~Thomson,
\newblock {Development of virtual patient models for permanent implant
  brachytherapy Monte Carlo dose calculations: interdependence of CT image
  artifact mitigation and tissue assignment},
\newblock Phys. Med. Biol. {\bf 60}, 6039 (2015).

\bibitem{TG43U1}
M.~J. Rivard, B.~M. Coursey, L.~A. DeWerd, W.~F. Hanson, M.~Saiful~Huq, G.~S.
  Ibbott, M.~G. Mitch, R.~Nath, and J.~F. Williamson,
\newblock {Update of {AAPM} {Task} {Group} {No}. 43 {Report}: {A} revised
  {AAPM} protocol for brachytherapy dose calculations},
\newblock Med. Phys. {\bf 31}, 633--674 (2004).

\bibitem{python}
G.~Van~Rossum and F.~L. Drake,
\newblock {\em {Python 3 Reference Manual}},
\newblock CreateSpace, Scotts Valley, CA, 2009.

\bibitem{numpy}
C.~R. Harris et~al.,
\newblock {Array programming with NumPy},
\newblock Nature {\bf 585}, 357--362 (2020).

\bibitem{Cl13}
L.~P. Coelho,
\newblock {Mahotas: {Open} source software for scriptable computer vision},
\newblock J. of Open Research Software {\bf 1}, e3 (2013).

\bibitem{Kr14}
R.~M. Kumar and S.~K,
\newblock {A Survey on Image Feature Descriptors},
\newblock Int. J. of Computer Science and Information Technologies {\bf 5}
  (2014).

\bibitem{ICRU58}
ICRU,
\newblock {ICRU 58 (dose and volume specification for reporting interstitial
  therapy), by International Commission on Radiation Units and Measurements},
\newblock J. of the ICRU  (1998).

\bibitem{Ti11}
T.~Major, G.~Fröhlich, and C.~Polgar,
\newblock {Assessment of dose homogeneity in conformal interstitial breast
  brachytherapy with special respect to {ICRU} recommendations},
\newblock J. of Cont. Brachytherapy {\bf 3}, 150--156 (2011).

\bibitem{scipy}
P.~Virtanen et~al.,
\newblock {{SciPy} 1.0: Fundamental Algorithms for Scientific Computing in
  Python},
\newblock Nature Methods {\bf 17}, 261--272 (2020).

\bibitem{Gh18}
A.~Ghila, B.~G. Fallone, and S.~Rathee,
\newblock {Technical Note: Experimental verification of EGSnrc calculated depth
  dose within a parallel magnetic field in a lung phantom},
\newblock Med. Phys. {\bf 45}, 5653--5658 (2018).

\bibitem{Em21}
M.~A. Ebert, S.~Gulliford, O.~Acosta, R.~de~Crevoisier, T.~McNutt, W.~D.
  Heemsbergen, M.~Witte, G.~Palma, T.~Rancati, and C.~Fiorino,
\newblock {Spatial descriptions of radiotherapy dose: normal tissue
  complication models and statistical associations},
\newblock Phys. Med. Biol. {\bf 66}, 12TR01 (2021).

\bibitem{Pl21}
L.~Placidi, E.~Gioscio, C.~Garibaldi, T.~Rancati, A.~Fanizzi, D.~Maestri,
  R.~Massafra, E.~Menghi, A.~Mirandola, G.~Reggiori, R.~Sghedoni, P.~Tamborra,
  S.~Comi, J.~Lenkowicz, L.~Boldrini, and M.~Avanzo,
\newblock {A Multicentre Evaluation of Dosiomics Features Reproducibility,
  Stability and Sensitivity.},
\newblock Cancers (Basel) {\bf 13}, 3835 (2021).

\bibitem{Zk15}
K.~Zourari, T.~Major, A.~Herein, V.~Peppa, C.~Polgár, and P.~Papagiannis,
\newblock {A retrospective dosimetric comparison of {TG43} and a commercially
  available {MBDCA} for an {APBI} brachytherapy patient cohort},
\newblock Physica Medica {\bf 31}, 669--676 (2015).

\end{thebibliography}




\bibliographystyle{bibStyle.bst}


\end{document}


\title{\bfseries \Large Supplementary material for Haralick texture feature analysis for Monte Carlo dose distributions of permanent implant prostate brachytherapy}

\author{}

\date{\small August 31, 2024}

\newgeometry{top=0in,bottom=1in,right=1in,left=1in}

\begin{titlepage}

\thispagestyle{fancy}
\maketitle

\end{titlepage}

\pagenumbering{arabic}
\setcounter{page}{1}
\pagestyle{fancy}

\section{Differential dose volume histograms for six patient subset}\label{SM:sec:dDVH}

\begin{figure}[H]
	\centering
	\includegraphics[height =\linewidth]{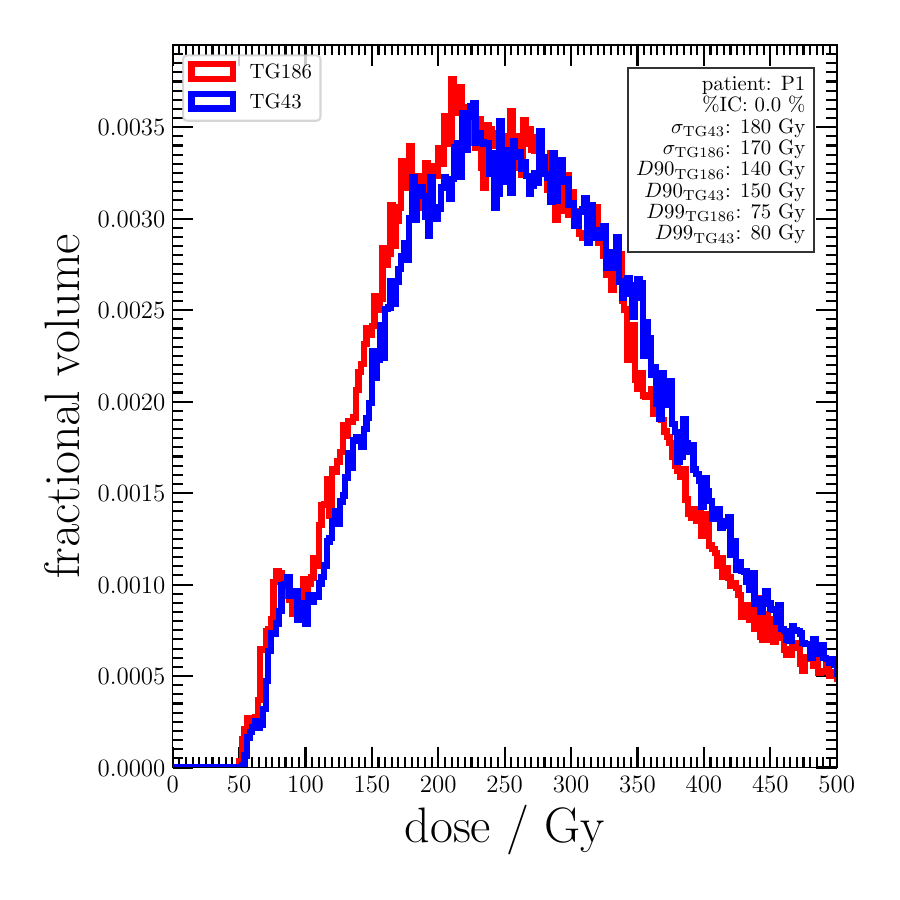}
	\caption[Differential dose-volume histograms using TG43 and TG186 conditions for P$_1$.]{Differential dose-volume histograms using TG43 and TG186 conditions for P$_1$. For both TG43 and TG186 conditions, measures of $D90$ and $D99$ (minimum absorbed dose received by hottest 90\% and 99\% of the target volume respectively) in conjunction with the standard deviation ($\sigma$) are presented.}
	\label{SM:fig:P1}
\end{figure}

\begin{figure}[H]
	\centering
	\includegraphics[height =\linewidth]{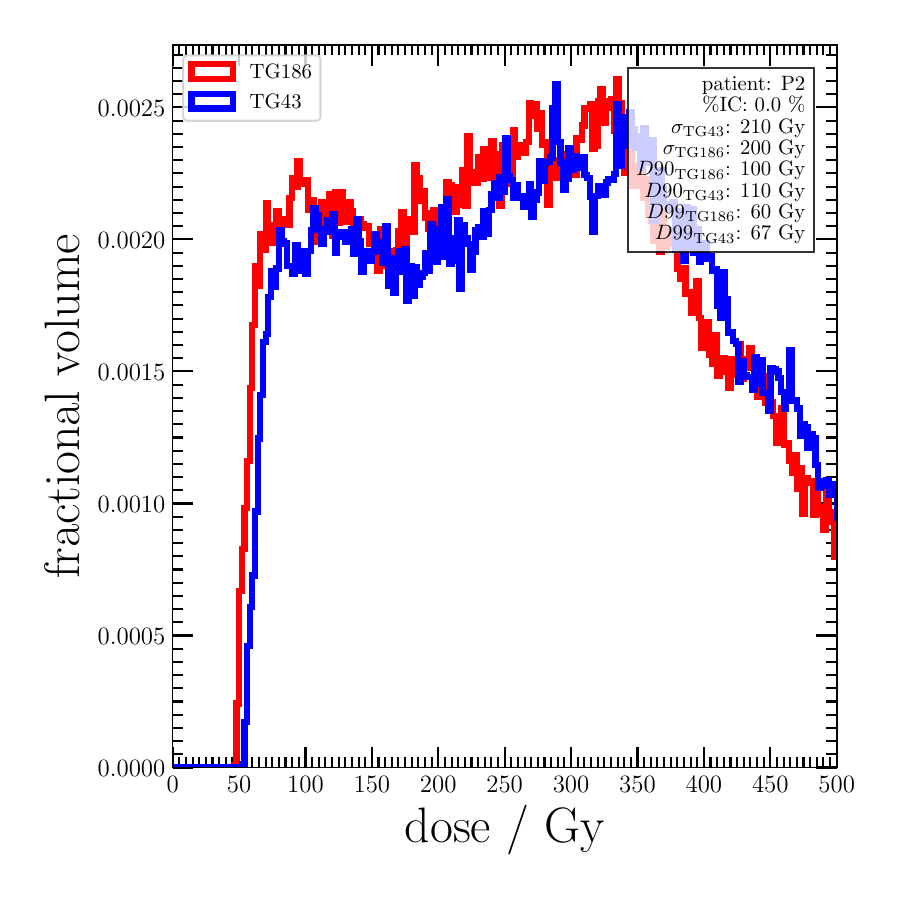}
	\caption[Differential dose-volume histograms using TG43 and TG186 conditions for P$_2$.]{Differential dose-volume histograms using TG43 and TG186 conditions for P$_2$. For both TG43 and TG186 conditions, measures of $D90$ and $D99$ (minimum absorbed dose received by hottest 90\% and 99\% of the target volume respectively) in conjunction with the standard deviation ($\sigma$) are presented.}
	\label{SM:fig:P2}
\end{figure}

\begin{figure}[H]
	\centering
	\includegraphics[height =\linewidth]{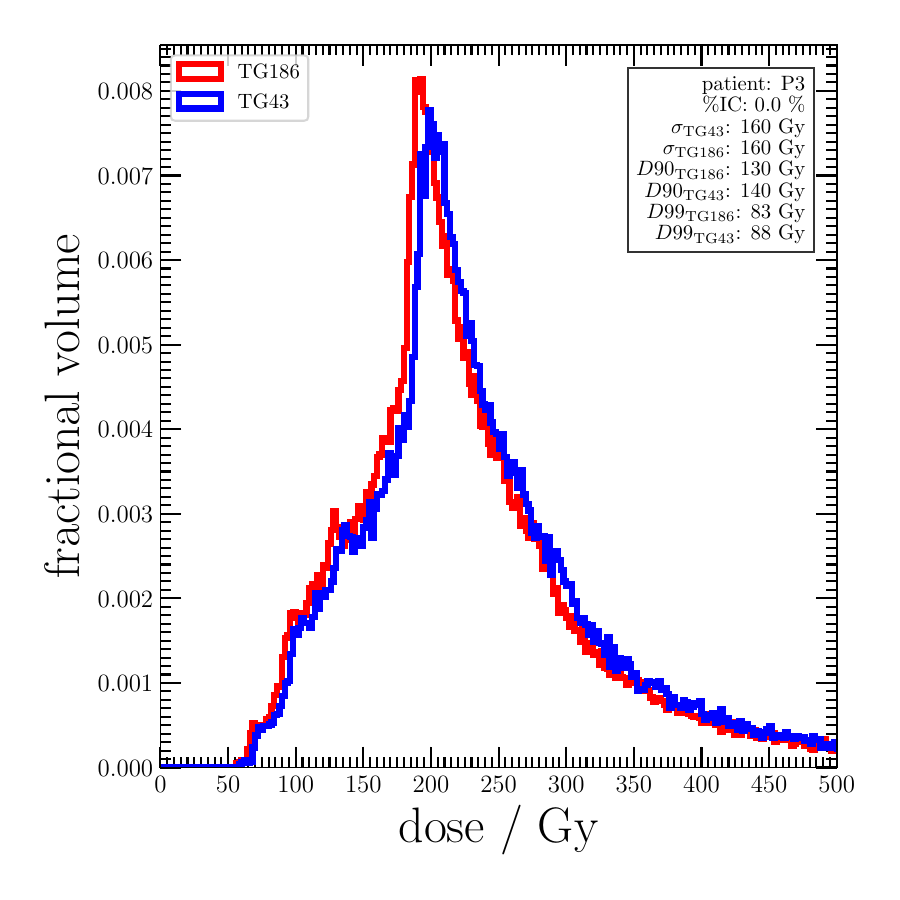}
		\caption[Differential dose-volume histograms using TG43 and TG186 conditions for P$_3$.]{Differential dose-volume histograms using TG43 and TG186 conditions for P$_3$. For both TG43 and TG186 conditions, measures of $D90$ and $D99$ (minimum absorbed dose received by hottest 90\% and 99\% of the target volume respectively) in conjunction with the standard deviation ($\sigma$) are presented.}
	\label{SM:fig:P3}
\end{figure}

\begin{figure}[H]
	\centering
	\includegraphics[height =\linewidth]{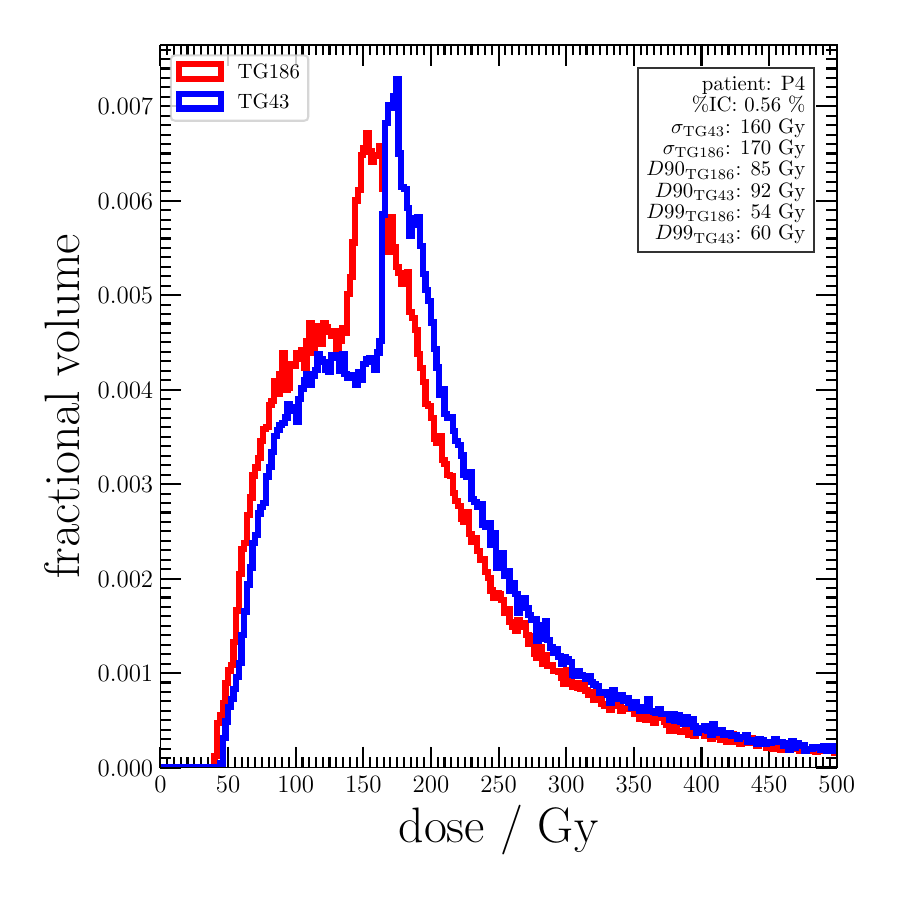}
	\caption[Differential dose-volume histograms using TG43 and TG186 conditions for P$_4$.]{Differential dose-volume histograms using TG43 and TG186 conditions for P$_4$. For both TG43 and TG186 conditions, measures of $D90$ and $D99$ (minimum absorbed dose received by hottest 90\% and 99\% of the target volume respectively) in conjunction with the standard deviation ($\sigma$) are presented.}
	\label{SM:fig:P4}
\end{figure}

\begin{figure}[H]
	\centering
	\includegraphics[height =\linewidth]{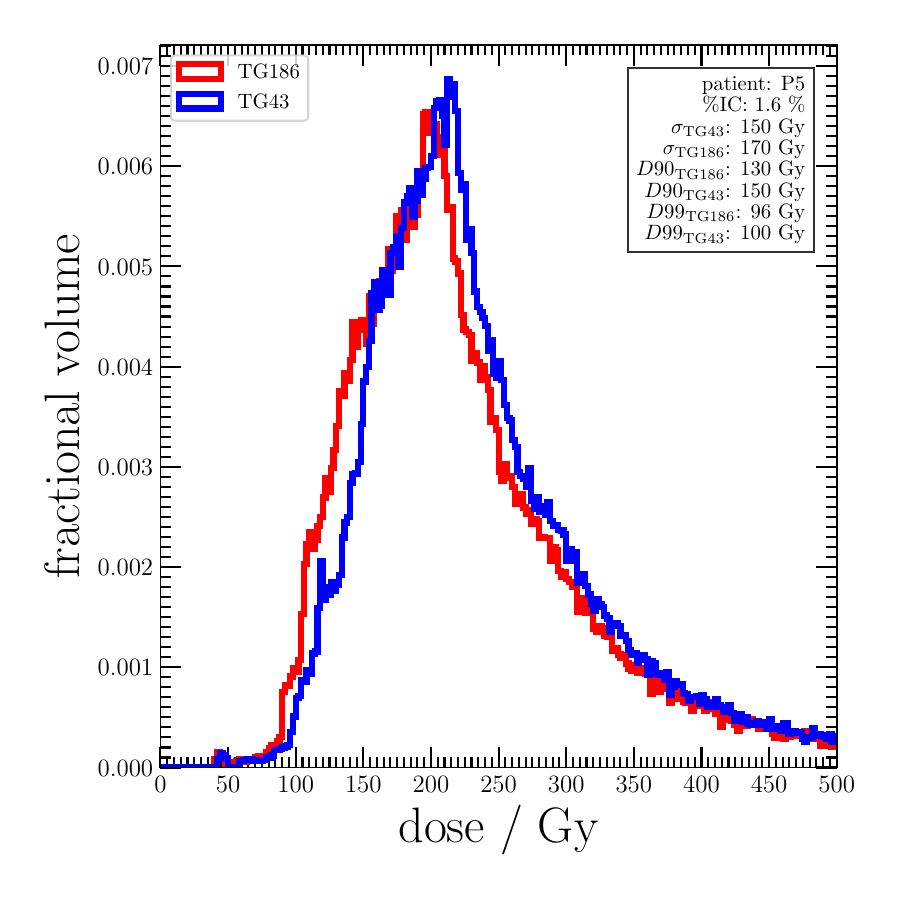}
		\caption[Differential dose-volume histograms using TG43 and TG186 conditions for P$_5$.]{Differential dose-volume histograms using TG43 and TG186 conditions for P$_5$. For both TG43 and TG186 conditions, measures of $D90$ and $D99$ (minimum absorbed dose received by hottest 90\% and 99\% of the target volume respectively) in conjunction with the standard deviation ($\sigma$) are presented.}
	\label{SM:fig:P5}
\end{figure}

\begin{figure}[H]
	\centering
	\includegraphics[height =\linewidth]{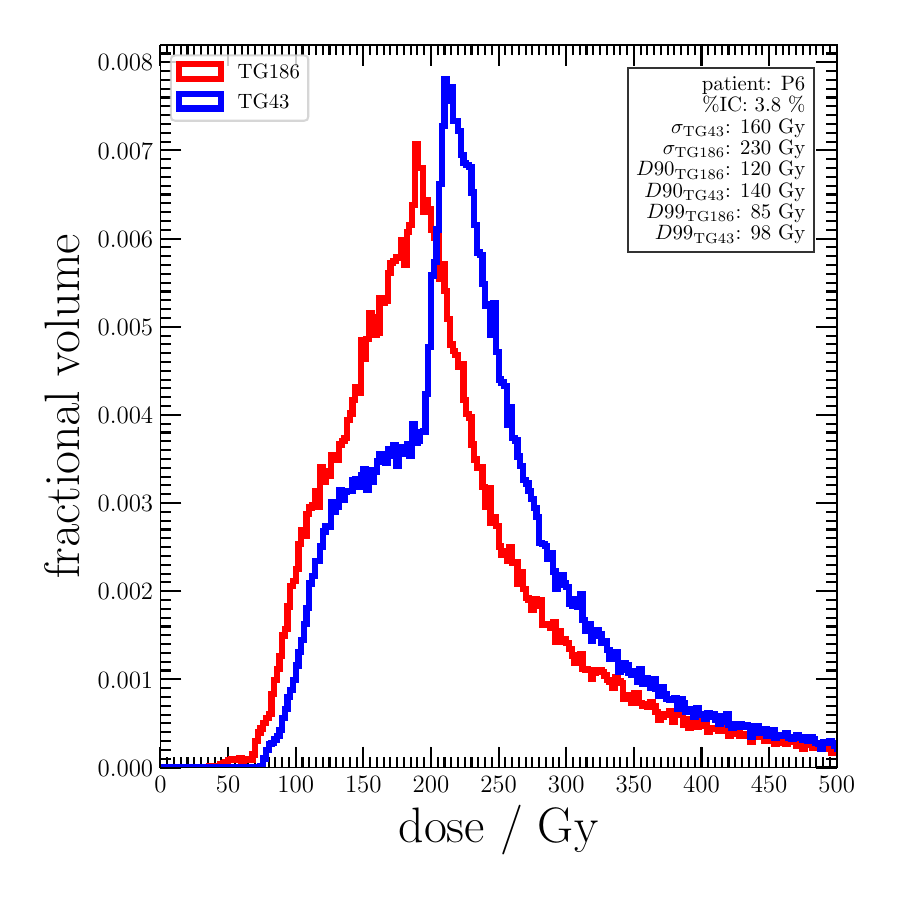}
		\caption[Differential dose-volume histograms using TG43 and TG186 conditions for P$_6$.]{Differential dose-volume histograms using TG43 and TG186 conditions for P$_6$. For both TG43 and TG186 conditions, measures of $D90$ and $D99$ (minimum absorbed dose received by hottest 90\% and 99\% of the target volume respectively) in conjunction with the standard deviation ($\sigma$) are presented.}
	\label{SM:fig:P6}
\end{figure}

\pagebreak
\section{Absorbed dose distributions for six patient subset}\label{SM:sec:doseDist}

\begin{figure}[H]
	\centering
\includegraphics[width=\linewidth]{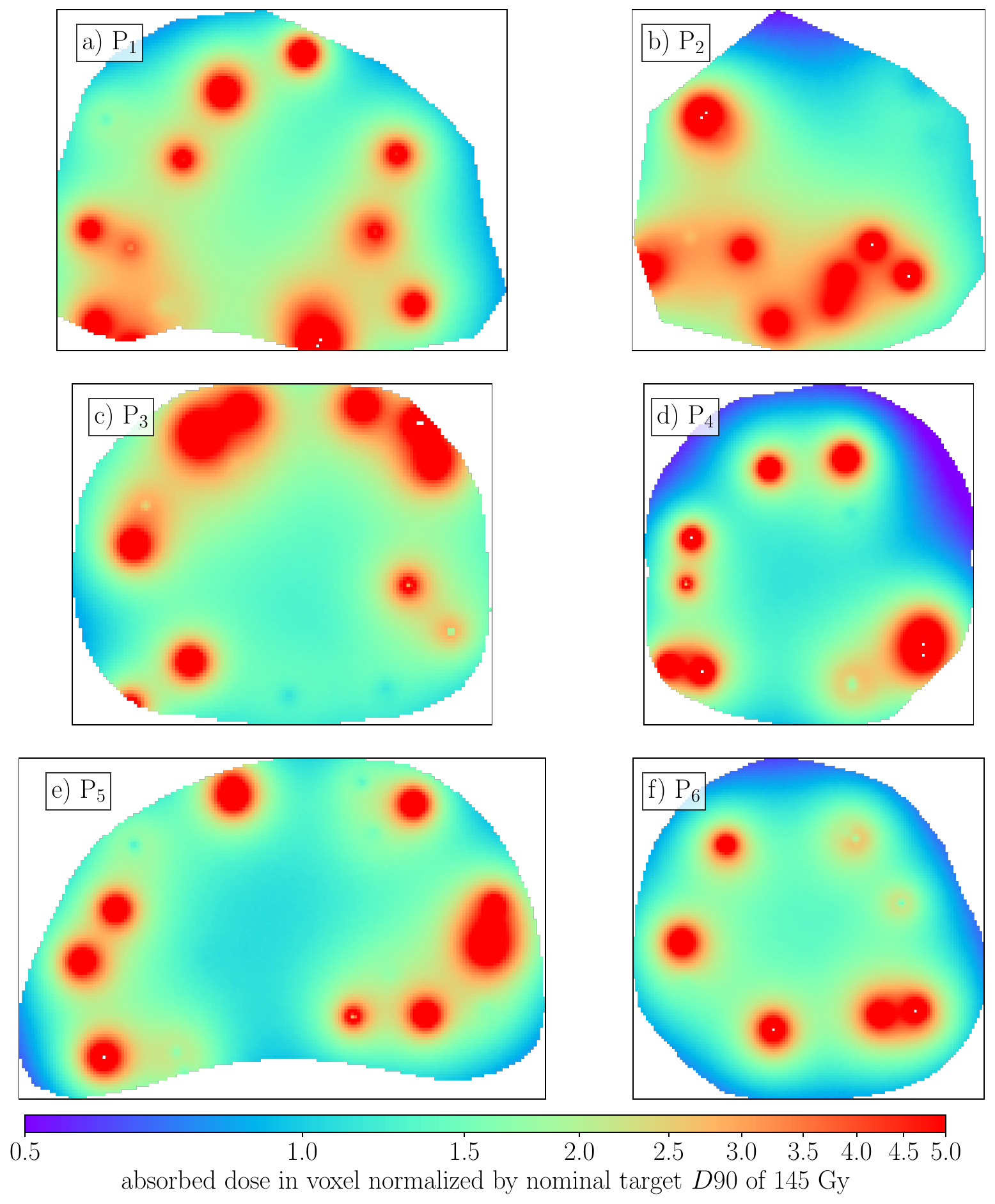}
	\caption[Patient specific absorbed dose distributions calculated using TG43 conditions]{Absorbed dose distributions calculated under TG43 conditions for six patients. For each distribution presented, the magnitude of absorbed dose in each voxel is normalized by the nominal target $D90$ of 145 Gy.}
	\label{fig:RawDoseTG43}
\end{figure}

\begin{figure}[H]
	\centering
\includegraphics[width=\linewidth]{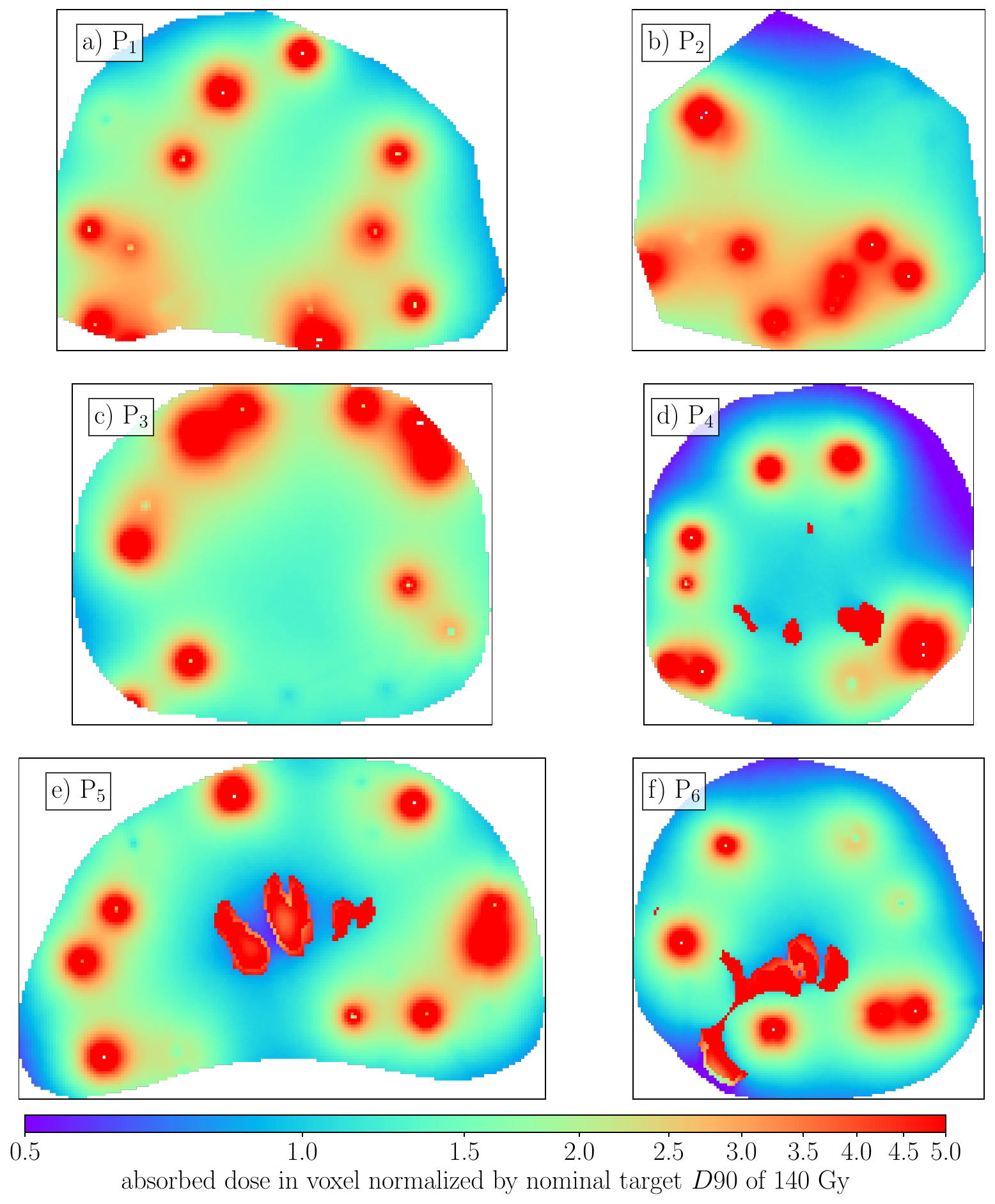}
	\caption[Patient specific absorbed dose distributions calculated using TG186 conditions]{Absorbed dose distributions calculated under TG186 conditions for six patients (P$_1$ - P$_6$) with (a,b,c) 0 \%IC, (d) 0.56 \%IC, (e) 1.6 \%IC, and (f) 3.8 \%IC. For each distribution presented, the magnitude of absorbed dose in each voxel is normalized by the nominal target $D90$ of 145 Gy. Intraprostatic calcifications are visible in (d), (e) and (f) as contiguous `red' regions corresponding to doses much larger than the prescription dose.}
	\label{fig:RawDose}
\end{figure}

\pagebreak
\section{Voxel-by-voxel difference in absorbed dose distributions calculated using TG43 and TG186 conditions}\label{SM:sec:difference}

\begin{figure}[H]
	\centering
\includegraphics[height =\linewidth]{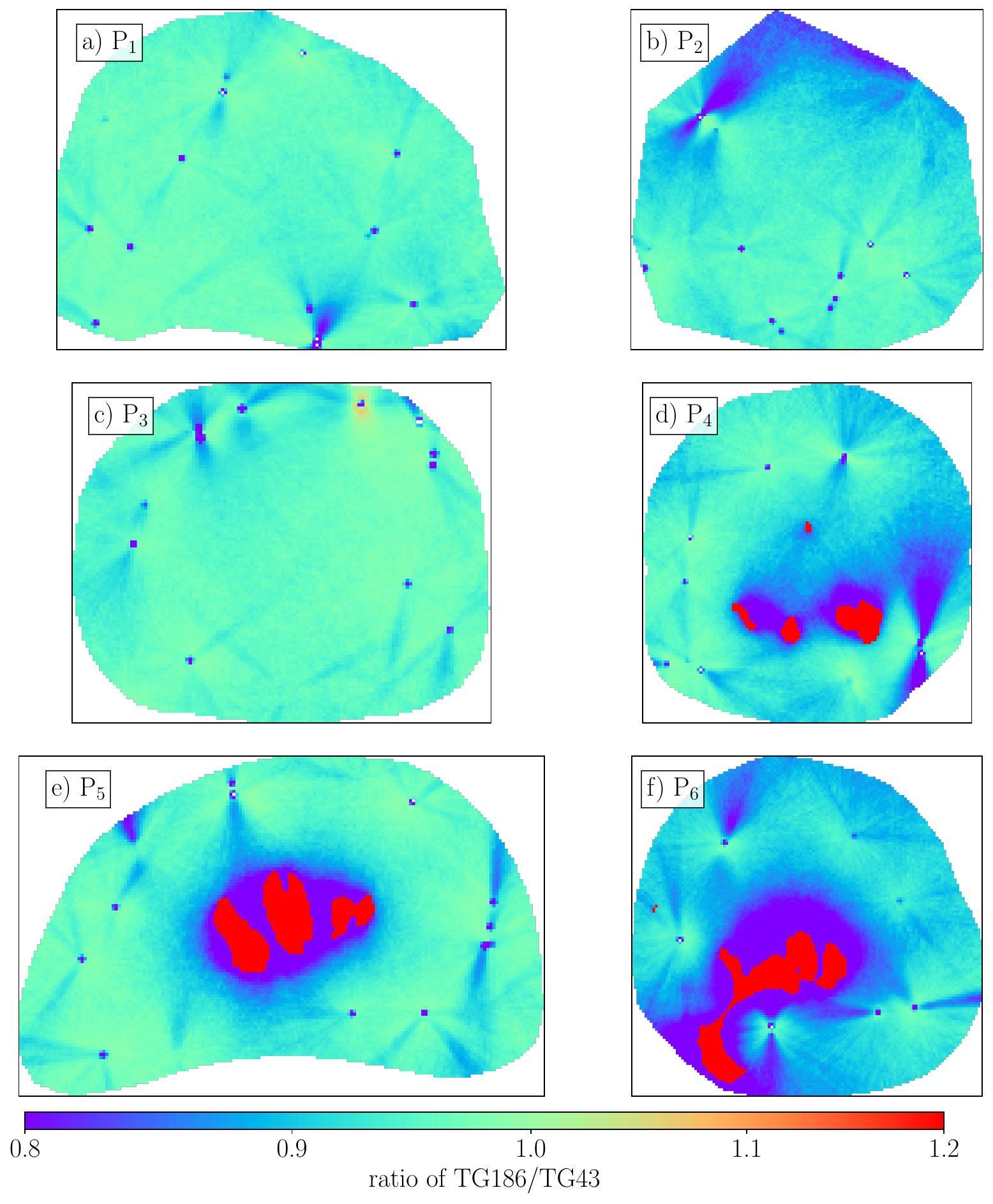}
	\caption[Voxel-by-voxel ratio between absorbed dose distributions calculated under TG186 and TG43 conditions for patients with varying levels of intraprostatic calcification (0-3.8 \%IC)]{Voxel-by-voxel ratio between absorbed dose distributions calculated under TG186 and TG43 conditions for patients with (a,b,c) 0, (d) 0.56, (e) 1.6, and (f) 3.8 \%IC.}
	\label{SM:fig:DoseDist}
\end{figure}

\pagebreak
\subsection{Textural measures extracted from patient specific 3D absorbed dose distributions}\label{SM:sec:RawFeatures}

\begin{figure}[H]
	\centering
\includegraphics[height =\linewidth]{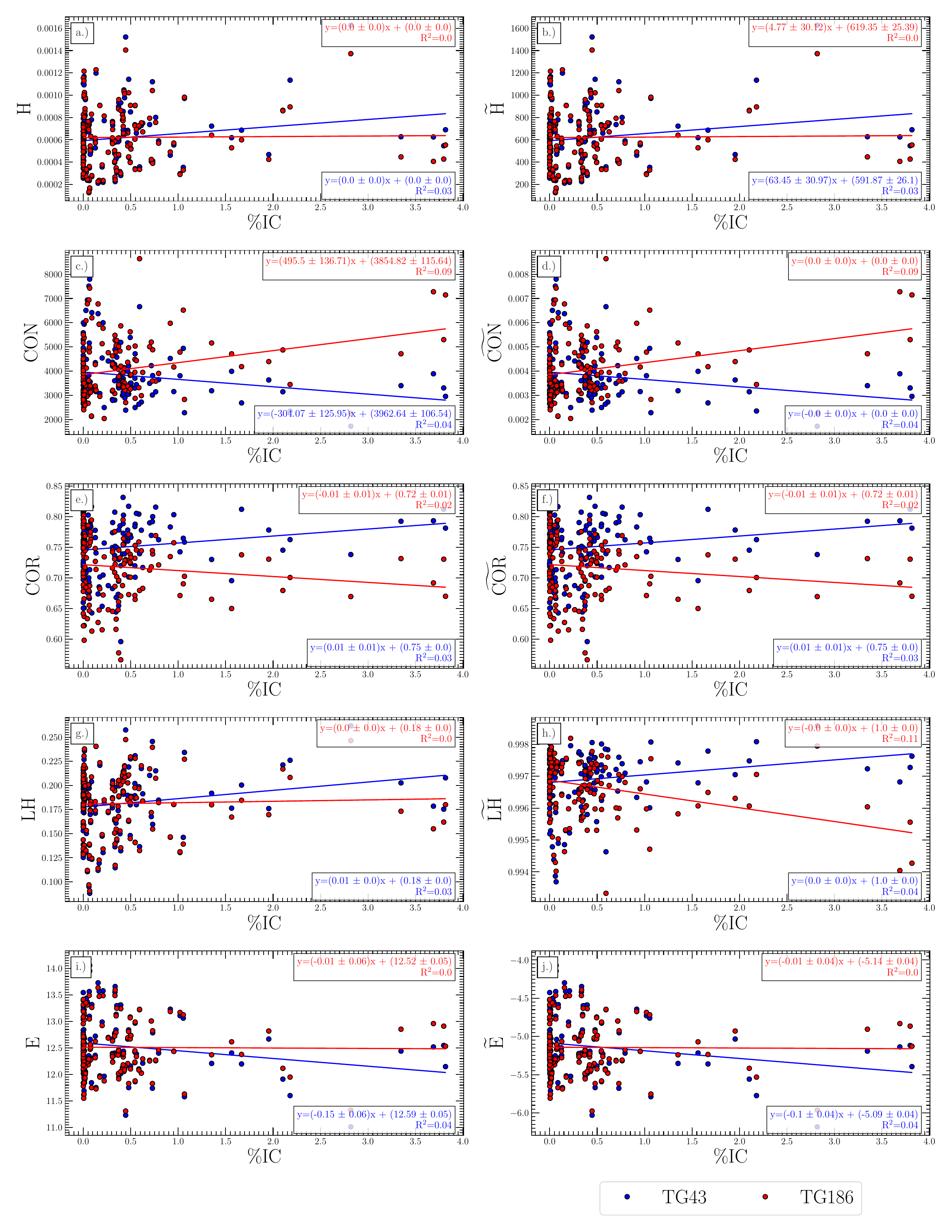}
	\caption[Textural features extracted from patient specific PIPB absorbed dose distributions using TG43 and TG186 conditions as a function of intraprostatic calcification percent by total prostate volume (\%IC) for original and modified textural measures.]{Textural features extracted from absorbed dose distributions using TG43 and TG186 conditions as a function of intraprostatic calcification percent by total prostate volume (\%IC) for (a,c,e,g,i) original and (b,d,f,h,j) modified textural measures.}
	\label{SM:fig:IC_diff}
\end{figure}

\begin{figure}[H]
	\centering
	\includegraphics[height =\linewidth]{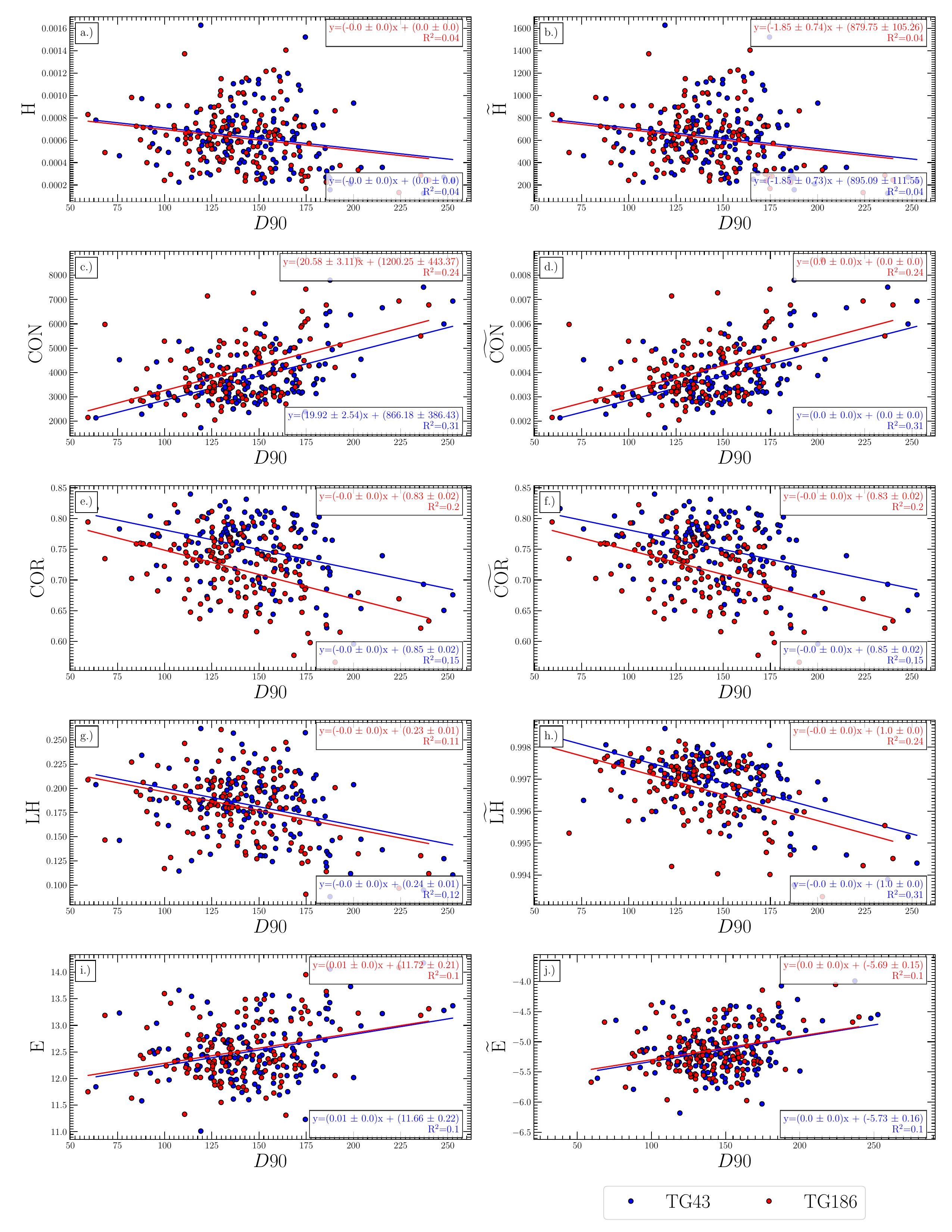}
	\caption[Textural features extracted from patient specific PIPB absorbed dose distributions using TG43 and TG186 conditions as a function of percent difference between measures of $D90$ for both original and modified textural measures.]{Textural features extracted from absorbed dose distributions using TG43 and TG186 conditions as a function of percent difference between measures of $D90$ (\DDninty{}) calculated under TG43 and TG186 conditions for (a,c,e,g,i) original and (b,d,f,h,j) modified textural measures.}
	\label{SM:fig:D90}
\end{figure}

\begin{figure}[H]
	\centering
	\includegraphics[height =\linewidth]{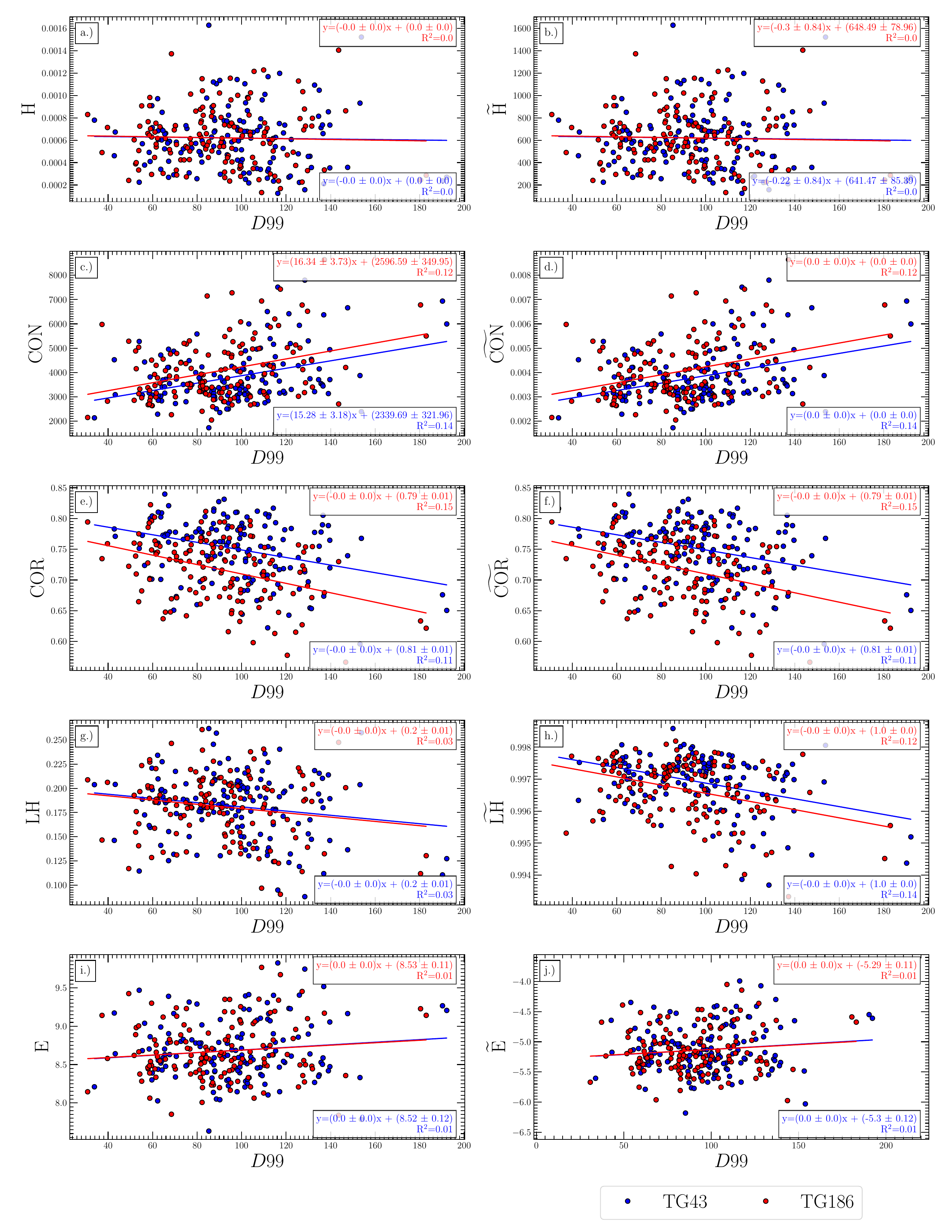}
	\caption[Textural features extracted from patient specific PIPB absorbed dose distributions using TG43 and TG186 conditions as a function of $D99$ for both original and modified textural measures.]{Textural features extracted from absorbed dose distributions using TG43 and TG186 conditions as a function of $D99$ calculated under TG43 and TG186 conditions for (a,c,e,g,i) original and (b,d,f,h,j) modified textural measures.}
	\label{SM:fig:D99}
\end{figure}

\pagebreak
\subsection{Relative Haralick measures for comparing TG43 and TG186 conditions in PIPB}\label{SM:sec:Relative}

\begin{figure}[H]
	\centering
\includegraphics[height =\linewidth]{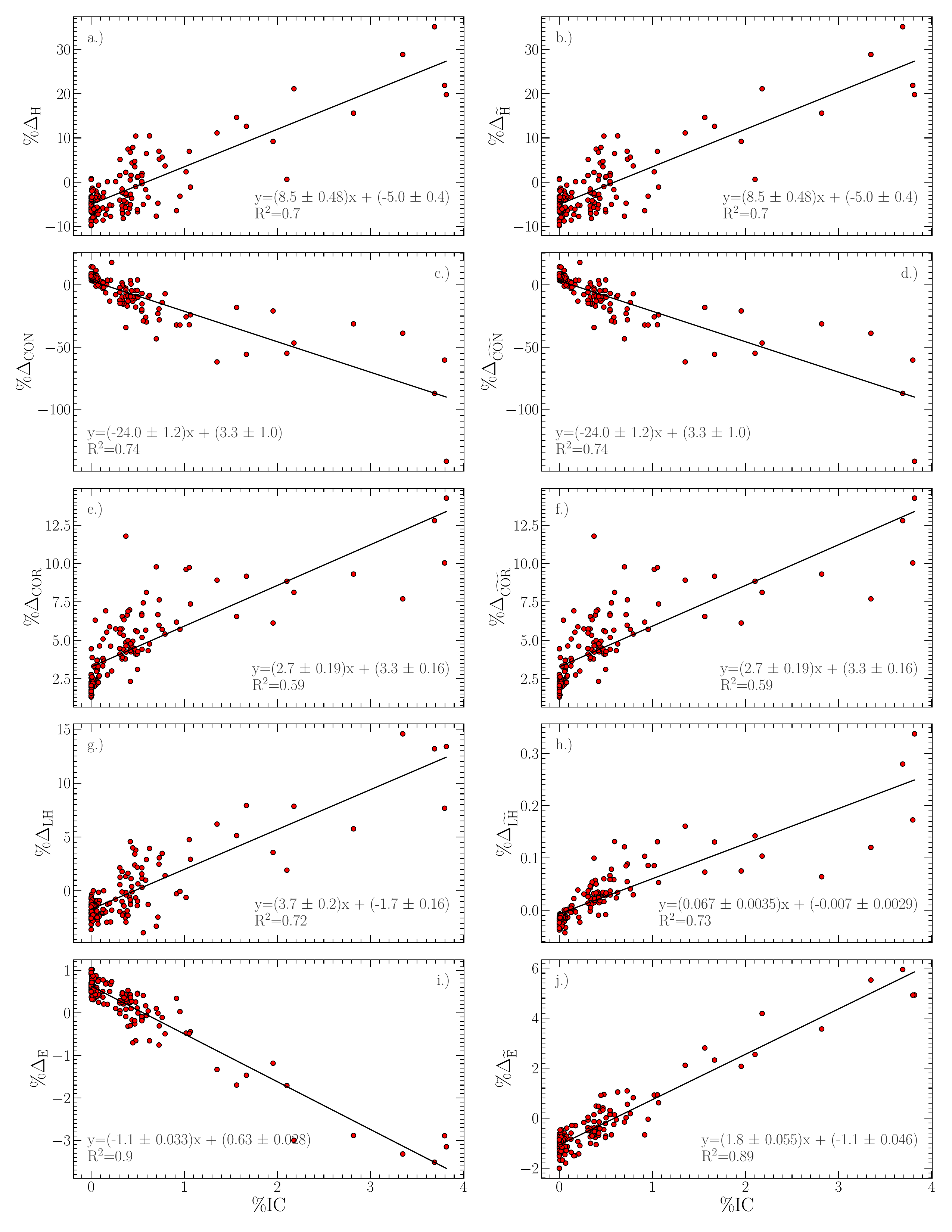}
	\caption[Percent difference between textural features extracted from patient specific PIPB absorbed dose distributions using TG43 and TG186 conditions as a function of intraprostatic calcification percent by total prostate volume (\%IC) for original and modified textural measures.]{Percent difference between textural features extracted from absorbed dose distributions using TG43 and TG186 conditions as a function of intraprostatic calcification percent by total prostate volume (\%IC) for (a,c,e,g,i) original and (b,d,f,h,j) modified textural measures.}
	\label{SM:fig:IC_diff}
\end{figure}

\begin{figure}[H]
	\centering
	\includegraphics[height =\linewidth]{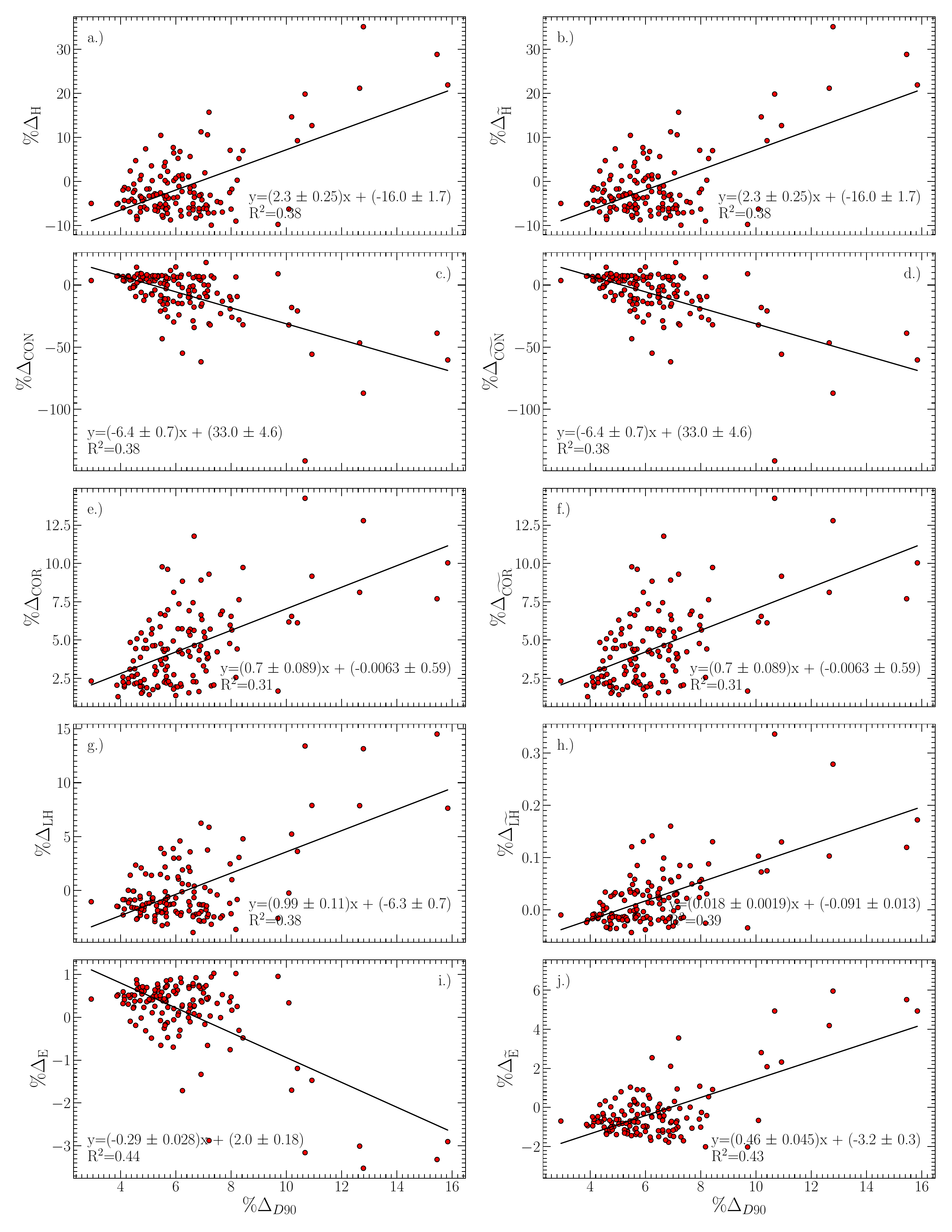}
	\caption[Percent difference between textural features extracted from patient specific PIPB absorbed dose distributions using TG43 and TG186 conditions as a function of $D90$ for original and modified textural measures.]{Percent difference between textural features extracted from dose distributions calculated under TG43 and TG186 conditions as a function of percent difference between measures of $D90$ calculated under TG43 and TG186 conditions for (a,c,e,g,i) original and (b,d,f,h,j) modified textural measures.}
	\label{SM:fig:D90_diff}
\end{figure}

\pagebreak
\section{Sensitivity of relative trends to quantization bin width} \label{SM:sec:QuantSens}

\begin{figure}[H]
	\centering
	\includegraphics[height =\linewidth]{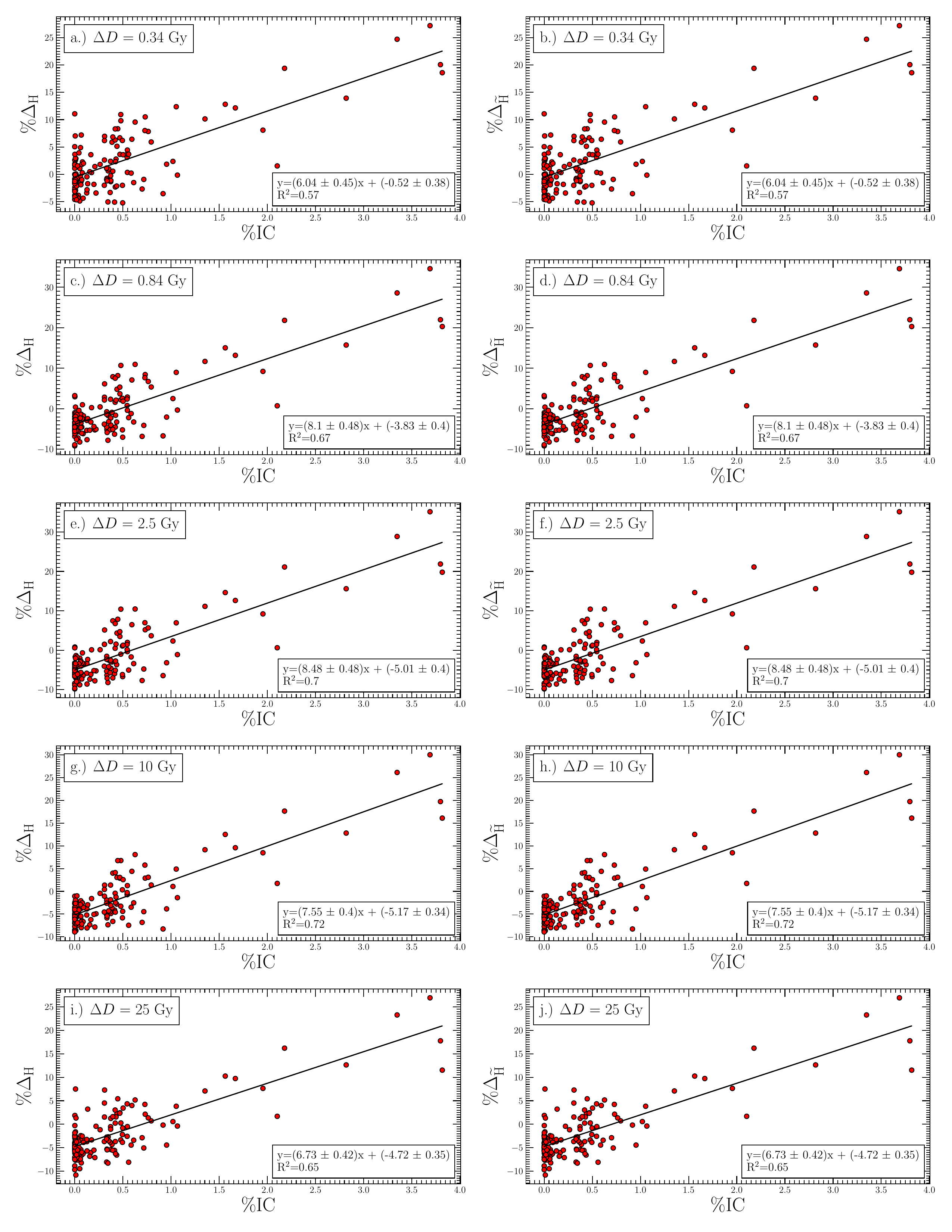}
	\caption[Homogeneity extracted from patient specific PIPB absorbed dose distributions with varying quantization level width ($\Delta D$: 0.3-25 Gy)]{Original and modified textural measures of homogeneity extracted from patient specific PIPB absorbed dose distributions with quantization level width ($\Delta D$) of (a,b) 0.3 Gy, (c,d) 0.8 Gy, (e,f) 2.5 Gy, (g,h) 10 Gy and (i,j) 25 Gy for (a,c,e,g,i) original and (b,d,f,h,j) modified textural features as a function of intraprostatic calcification by volume (\%IC).}
	\label{SM:fig:SensitivityFixedFeatureHom}
\end{figure}

\begin{figure}[H]
	\centering
	\includegraphics[height =\linewidth]{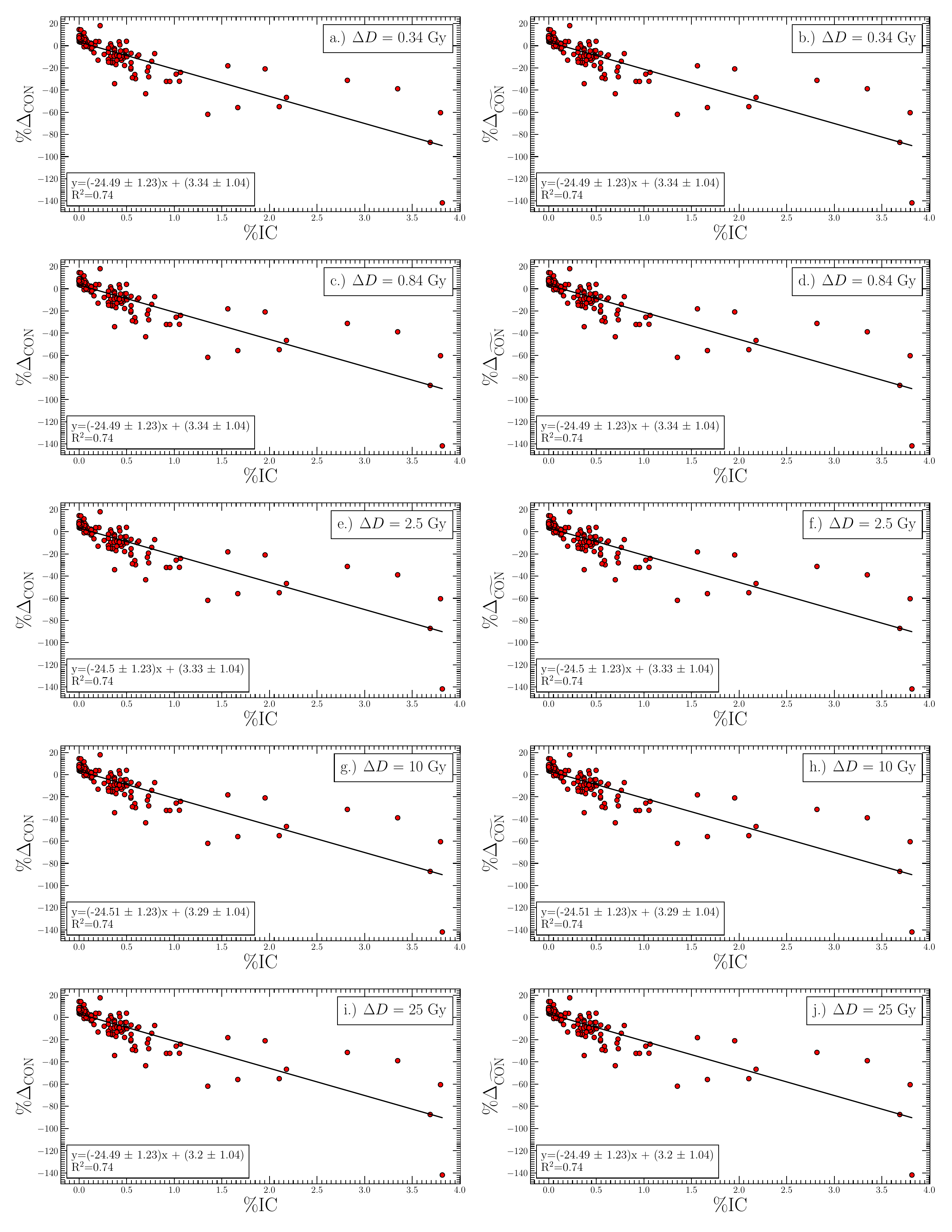}
	\caption[Contrast extracted from patient specific PIPB absorbed dose distributions with varying quantization level width ($\Delta D$: 0.3-25 Gy)]{Original and modified textural measures of contrast extracted from patient specific PIPB absorbed dose distributions with quantization level width ($\Delta D$) of (a,b) 0.3 Gy, (c,d) 0.8 Gy, (e,f) 2.5 Gy, (g,h) 10 Gy and (i,j) 25 Gy for (a,c,e,g,i) original and (b,d,f,h,j) modified textural features as a function of intraprostatic calcification by volume (\%IC).}
	\label{SM:fig:SensitivityFixedFeatureCon}
\end{figure}

\begin{figure}[H]
	\centering
	\includegraphics[height =\linewidth]{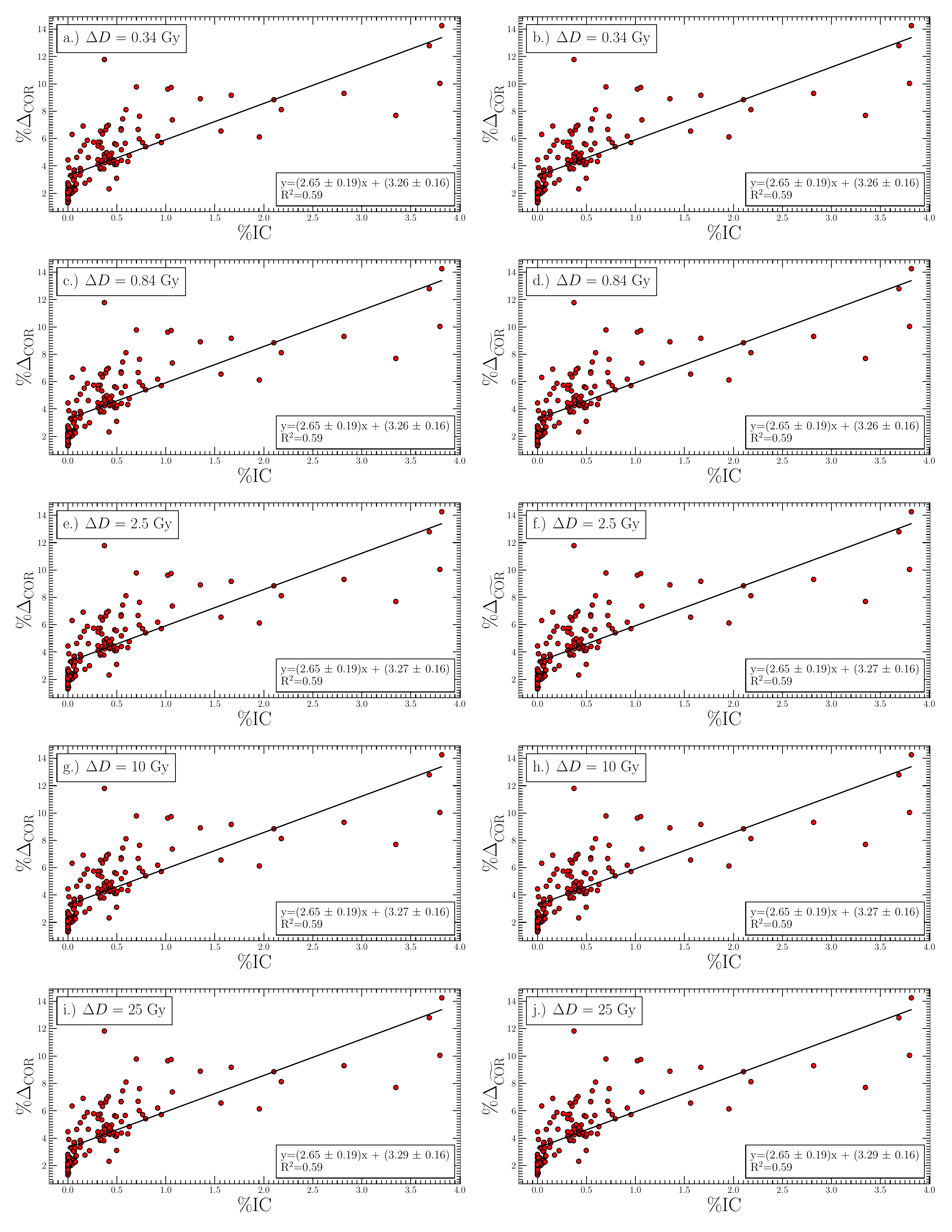}
	\caption[Correlation extracted from patient specific PIPB absorbed dose distributions with varying quantization level width ($\Delta D$: 0.3-25 Gy)]{Original and modified textural measures of correlation extracted from patient specific PIPB absorbed dose distributions with quantization level width ($\Delta D$) of (a,b) 0.3 Gy, (c,d) 0.8 Gy, (e,f) 2.5 Gy, (g,h) 10 Gy and (i,j) 25 Gy for (a,c,e,g,i) original and (b,d,f,h,j) modified textural features as a function of intraprostatic calcification by volume (\%IC).}
	\label{SM:fig:SensitivityFixedFeatureCor}
\end{figure}

\begin{figure}[H]
	\centering
	\includegraphics[height =\linewidth]{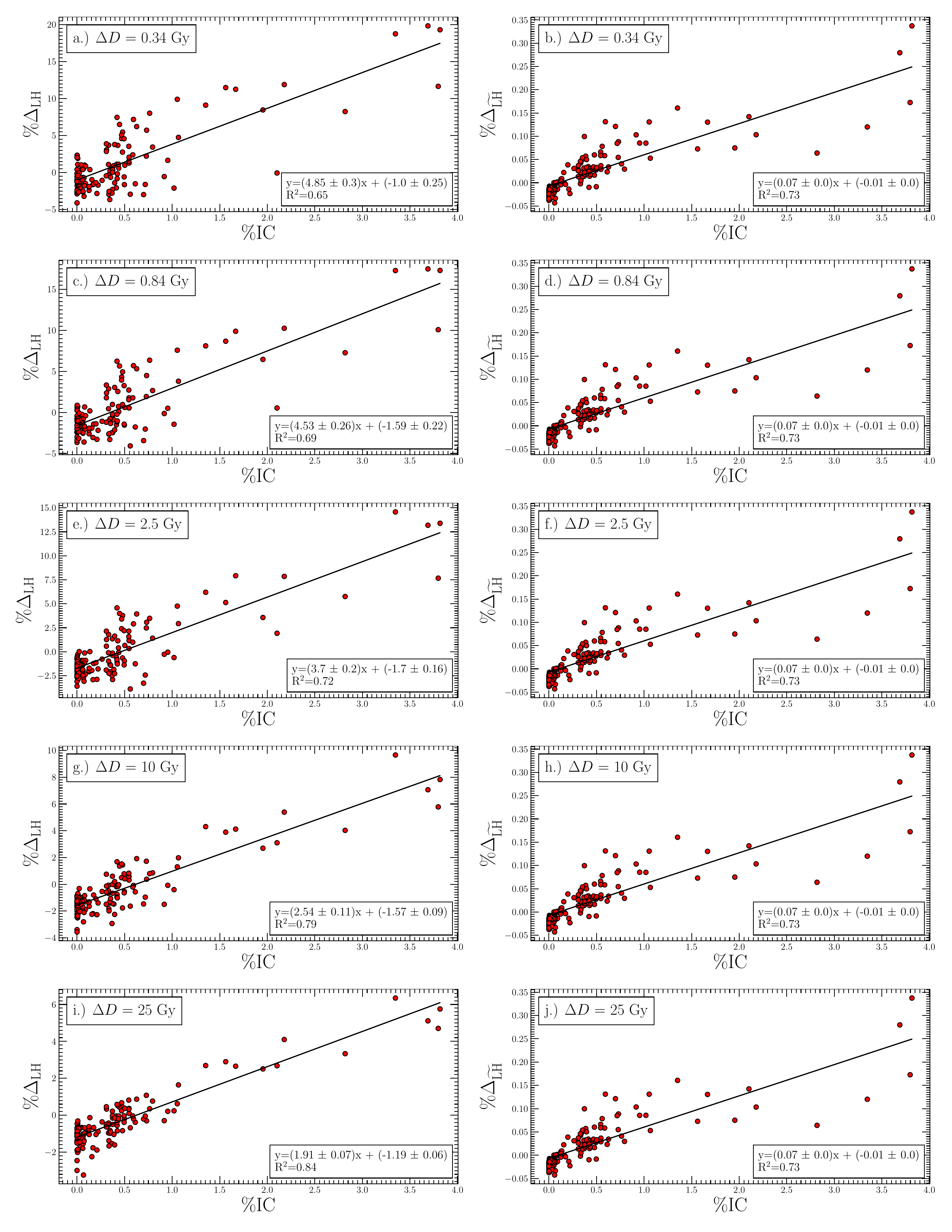}
	\caption[Local homogeneity extracted from patient specific PIPB absorbed dose distributions with varying quantization level width ($\Delta D$: 0.3-25 Gy)]{Original and modified textural measures of local homogeneity extracted from patient specific PIPB absorbed dose distributions with quantization level width ($\Delta D$) of (a,b) 0.3 Gy, (c,d) 0.8 Gy, (e,f) 2.5 Gy, (g,h) 10 Gy and (i,j) 25 Gy for (a,c,e,g,i) original and (b,d,f,h,j) modified textural features as a function of intraprostatic calcification by volume (\%IC).}
	\label{SM:fig:SensitivityFixedFeatureLH}
\end{figure}

\begin{figure}[H]
	\centering
	\includegraphics[height =\linewidth]{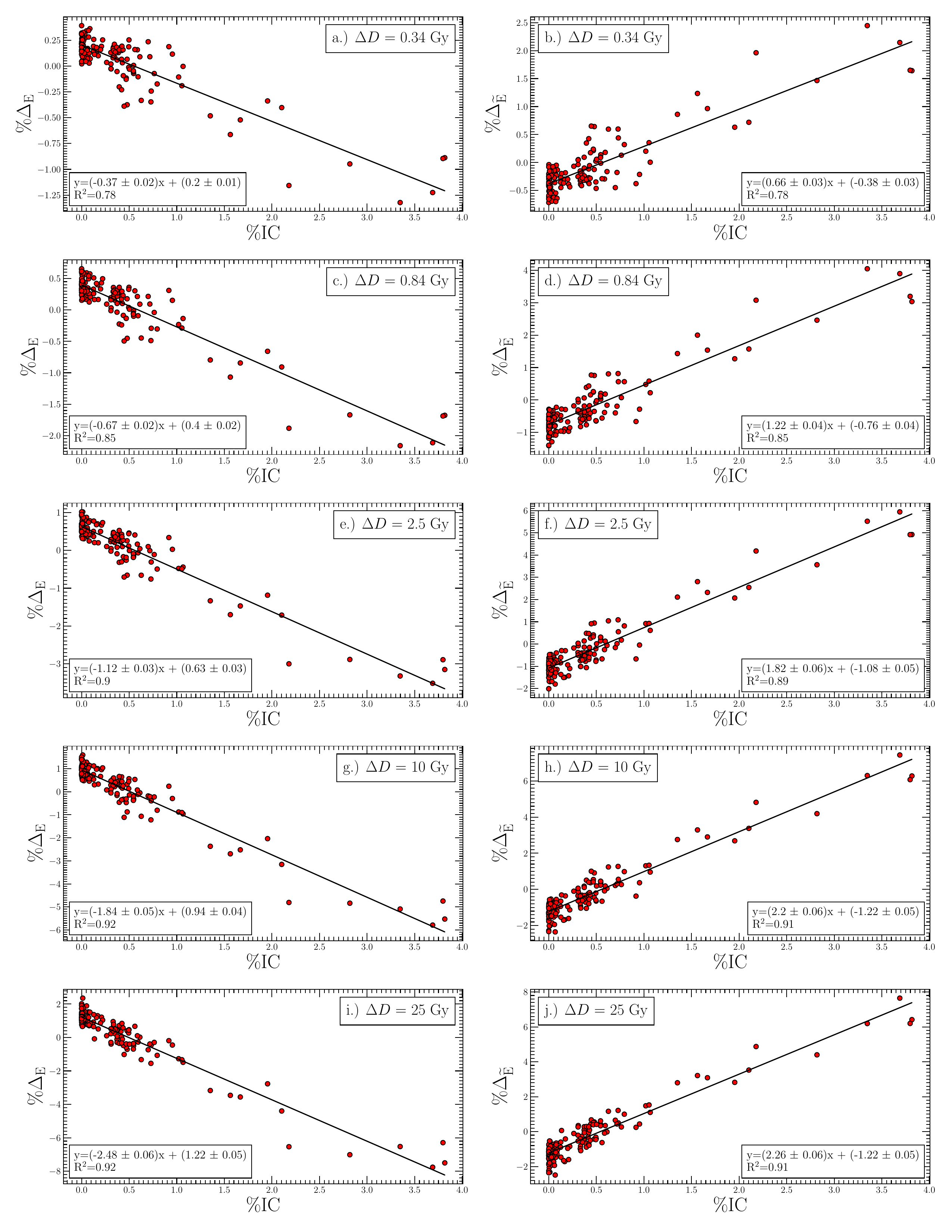}
	\caption[Entropy extracted from patient specific PIPB absorbed dose distributions with varying quantization level width ($\Delta D$: 0.3-25 Gy)]{Original and modified textural measures of entropy extracted from patient specific PIPB absorbed dose distributions with quantization level width ($\Delta D$) of (a,b) 0.3 Gy, (c,d) 0.8 Gy, (e,f) 2.5 Gy, (g,h) 10 Gy and (i,j) 25 Gy for (a,c,e,g,i) original and (b,d,f,h,j) modified textural features as a function of intraprostatic calcification by volume (\%IC).}
	\label{SM:fig:SensitivityFixedFeatureE}
\end{figure}

\pagebreak
\section{Comparison of original and modified textural measures}\label{SM:sec:analyticalCompare}

\subsection*{Original Haralick measures}
\begin{align}
\textrm{H} &= \sum_{i=1}^{N}  \sum_{j=1}^{N} p_{\theta}(i,j)^2 \\
\textrm{CON} &= \sum_{i=1}^{N} \sum_{j=1}^{N} (i - j)^2 p_{\theta}(i,j) \\
\textrm{COR} &= \sum_{i=1}^{N}\sum_{j=1}^{N} \left( \frac{i - \mu_{x}}{\sigma_{x}} \right) \left(\frac{j - \mu_{y}}{\sigma_{y}} \right)p_{\theta}(i,j) \\ 
\textrm{LH} &= \sum_{i=1}^{N}  \sum_{j=1}^{N} \frac{1}{1+(i - j)^2}p_{\theta}(i,j) \\
\textrm{E} &= - \sum_{i=1}^{N}  \sum_{j=1}^{N}p_{\theta}(i,j)\log(p_{\theta}(i,j))
\end{align}
where $p_{x}(i)~=~\sum_{j=1}^{N} p_{\theta}(i,j)$, $p_{y}(j)~=~\sum_{i=1}^{N} p_{\theta}(i,j)$, $\mu_x~=~\sum_{i=1}^{N}i\cdot~p_{x}(i)$, $\mu_{y}~=~\sum_{j=1}^{N}j\cdot~p_{y}(j)$, $\sigma_{x}^2~=~\sum_{i=1}^{N}(i-\mu_x)^2\cdot~p_{x}(i)$, $\sigma_{y}^2~=~\sum_{j=1}^{N}(j-\mu_{y})^2\cdot ~p_{y}(j)$.

\subsection*{Modified Haralick measures}
\begin{align}
\widetilde{\mathrm{H}} &= \sum_{i=1}^{N} \sum_{j=1}^{N} \widetilde{p}_{\theta}(i,j)^2 \Delta_{ij} \\
\widetilde{\mathrm{CON}} &= \sum_{i=1}^{N}\sum_{j=1}^{N} \left(\frac{i}{N} - \frac{j}{N}\right)^2 \widetilde{p}_{\theta}(i,j) \Delta_{ij} \\
\widetilde{\mathrm{COR}} &= \sum_{i=1}^{N}\sum_{j=1}^{N} \left( \frac{\frac{i}{N} - \widetilde{\mu}_{x}}{\widetilde{\sigma}_{x}} \right) \left( \frac{\frac{j}{N} - \widetilde{\mu}_{y}}{\widetilde{\sigma}_{y}} \right) \widetilde{p}_{\theta}(i,j) \Delta_{ij} \\ 
\widetilde{\mathrm{LH}} &= \sum_{i=1}^{N} \sum_{j=1}^{N} \frac{1}{1+\left(\frac{i}{N} - \frac{j}{N}\right)^2}\widetilde{p}_{\theta}(i,j)\Delta_{ij} \\
\widetilde{\mathrm{E}} &= - \sum_{i=1}^{N} \sum_{j=1}^{N}\widetilde{p}_{\theta}(i,j)\log(\widetilde{p}_{\theta}(i,j))\Delta_{ij}
\end{align}

\noindent
where $\Delta~=~\frac{1}{N}$, $\Delta_{ij}~=~\frac{1}{N^2}$, $\widetilde{p}_{x}(i)~=~\sum_{j=1}^{N}\widetilde{p}_{\theta}(i,j) \Delta$, $\widetilde{p}_{y}(j)~=~\sum_{i=1}^{N}\widetilde{p}_{\theta}(i,j) \Delta$, $\widetilde{\mu} _x~=~\sum_{i=1}^{N}\frac{i}{N}\cdot\widetilde{p}_{x}(i)$, $\widetilde{\mu} _{y}~=~\sum_{j=1}^{N}\frac{j}{N}\cdot \widetilde{p}_{y}(j)$, $\widetilde{\sigma}_{x}^2~=~\sum_{i=1}^{N} (\frac{i}{N}-\widetilde{\mu} _x)^2\cdot\widetilde{p}_{x}(i)$, $\widetilde{\sigma}_{y}^2~=~\sum_{j=1}^{N}(\frac{j}{N}-\widetilde{\mu} _{y})^2\cdot \widetilde{p}_{y}(j)$.

\subsection{Homogeneity}
\begin{align}
\widetilde{\mathrm{H}} &= \sum_{i=1}^{N} \sum_{j=1}^{N} \widetilde{p}_{\theta}(i,j)^2 \Delta_{ij} \nonumber \\
&= \sum_{i=1}^{N} \sum_{j=1}^{N} \left(  p_{\theta}(i,j)\cdot N^2 \right)^2 \left( \frac{1}{N^2} \right) \nonumber \\
&= \sum_{i=1}^{N} \sum_{j=1}^{N}  p_{\theta}(i,j)^2 N^2 \nonumber \\
&= N^2 \cdot \sum_{i=1}^{N} \sum_{j=1}^{N} p_{\theta}(i,j)^2 \nonumber \\
\widetilde{\mathrm{H}} &= N^2 \cdot \mathrm{H}
\end{align}

\subsection{Contrast}
\begin{align}
\widetilde{\mathrm{CON}} &= \sum_{i=1}^{N}\sum_{j=1}^{N} \left(\frac{i}{N} - \frac{j}{N}\right)^2 \widetilde{p}_{\theta}(i,j) \Delta_{ij} \nonumber \\
&= \sum_{i=1}^{N}\sum_{j=1}^{N}  \left(\frac{1}{N^2} \cdot (i - j)^2\right)  \left(  p_{\theta}(i,j)\cdot N^2 \right) \left( \frac{1}{N^2} \right) \nonumber \\
&= \frac{1}{N^2} \cdot \sum_{i=1}^{N}\sum_{j=1}^{N} (i - j)^2 p_{\theta}(i,j) \nonumber \\
\widetilde{\mathrm{CON}} &= \frac{1}{N^2} \cdot \textrm{CON}
\end{align}

\subsection{Correlation}\label{SM:correlation}
\noindent 
Using the following relationships: 
\begin{align}
\widetilde{\mu} _x &= \frac{\mu _x}{N} \nonumber \\
\widetilde{\sigma}_{x}^2 &=\frac{\sigma_{x}^2}{N^2} \nonumber \\
\widetilde{\mu} _y &= \frac{\mu _y}{N} \nonumber \\
\widetilde{\sigma}_{y}^2 &=\frac{\sigma_{y}^2}{N^2} \nonumber
\end{align}

\noindent 
modified correlation may be written in terms of original correlation:
\begin{align}
\widetilde{\mathrm{COR}} &= \sum_{i=1}^{N}\sum_{j=1}^{N} \left( \frac{\frac{i}{N} - \widetilde{\mu}_{x}}{\widetilde{\sigma}_{x}} \right) \left( \frac{\frac{j}{N} - \widetilde{\mu}_{y}}{\widetilde{\sigma}_{y}} \right) \widetilde{p}_{\theta}(i,j) \Delta_{ij} \nonumber \\
&= \sum_{i=1}^{N}\sum_{j=1}^{N} \left( \frac{\frac{i}{N} - \left( \frac{\mu _x}{N} \right)}{\frac{\sigma_{x}}{N}} \right) \left( \frac{\frac{j}{N} - \left( \frac{\mu _y}{N} \right)}{\frac{\sigma_{y}}{N}} \right) \left( p_{\theta}(i,j)\cdot N^2 \right) \left( \frac{1}{N^2} \right) \nonumber \\
&= \sum_{i=1}^{N}\sum_{j=1}^{N} \left( \frac{i - \mu_{x}}{\sigma_{x}} \right) \left(\frac{j - \mu_{y}}{\sigma_{y}} \right)p_{\theta}(i,j) \\ 
\widetilde{\mathrm{COR}} &= \textrm{COR} 
\end{align}

\subsection{Local Homogeneity}

\begin{align}
\widetilde{\mathrm{LH}} &= \sum_{i=1}^{N} \sum_{j=1}^{N} \frac{1}{1+\left(\frac{i}{N} - \frac{j}{N}\right)^2}\widetilde{p}_{\theta}(i,j)\Delta_{ij}
\end{align}

The leading term within the summation $\left( \frac{1}{1+\left(\frac{i}{N} - \frac{j}{N}\right)^2} \right)$ cannot be factored in a way to write the modified calculation of local homogeneity in terms of a simple scaling/translation of the original measure.

\subsection{Entropy}\label{SM:sec:Entropy}


\begin{align}
\widetilde{\mathrm{E}} &= - \sum_{i=1}^{N} \sum_{j=1}^{N}\widetilde{p}_{\theta}(i,j)\log(\widetilde{p}_{\theta}(i,j))\Delta_{ij} \nonumber \\ 
 &= - \sum_{i=1}^{N} \sum_{j=1}^{N}  \left(  p_{\theta}(i,j)\cdot N^2 \right) \log  \left(  p_{\theta}(i,j)\cdot N^2 \right) \left( \frac{1}{N^2} \right) \nonumber \\
  &= - \sum_{i=1}^{N} \sum_{j=1}^{N}  p_{\theta}(i,j) \left( \log(p_{\theta}(i,j)) + \log(N^2) \right) \nonumber \\
  &=  - \sum_{i=1}^{N}  \sum_{j=1}^{N}p_{\theta}(i,j) \log(p_{\theta}(i,j)) - \sum_{i=1}^{N} \sum_{j=1}^{N} p_{\theta}(i,j) \log(N^2) \nonumber \\
  &= \textrm{E} - \log(N^2) \cdot \underbrace{\sum_{i=1}^{N} \sum_{j=1}^{N} p_{\theta}(i,j)}_\text{=1}  \nonumber \\
 \widetilde{\mathrm{E}} &= \textrm{E} - 2\log(N)
\end{align}

\noindent
Considering case for calculations of E and $\widetilde{\mathrm{E}}$ based on $\log$ with bases of 2 and $e$ respectively:

\begin{align}
\widetilde{\mathrm{E}} &= - \sum_{i=1}^{N} \sum_{j=1}^{N}\widetilde{p}_{\theta}(i,j)\log_{e}(\widetilde{p}_{\theta}(i,j))\Delta_{ij} \nonumber \\ 
 &= - \sum_{i=1}^{N} \sum_{j=1}^{N}  \left(  p_{\theta}(i,j)\cdot N^2 \right) \log_{e}  \left(  p_{\theta}(i,j)\cdot N^2 \right) \left( \frac{1}{N^2} \right) \nonumber \\
 &= - \sum_{i=1}^{N} \sum_{j=1}^{N}  p_{\theta}(i,j) \left( \log_{e}(p_{\theta}(i,j)) + \log_{e}(N^2) \right) \nonumber \\
 &=  - \sum_{i=1}^{N}  \sum_{j=1}^{N}p_{\theta}(i,j) \log_{e}(p_{\theta}(i,j)) -  \underbrace{\sum_{i=1}^{N} \sum_{j=1}^{N} p_{\theta}(i,j) \log_{e}(N^2)}_\text{$=2\log_{e}(N)$} \nonumber \\
 &= - \sum_{i=1}^{N}  \sum_{j=1}^{N}p_{\theta}(i,j) \log_{e}(p_{\theta}(i,j)) - 2\log_{e}(N) \nonumber \\
 &= - \log_{e} (2) \sum_{i=1}^{N}  \sum_{j=1}^{N}p_{\theta}(i,j) \frac{\log_{e}(p_{\theta}(i,j))}{ \log_{e} (2)}  - 2\log_{e}(N) \nonumber \\
&= - \log_{e} (2) \sum_{i=1}^{N}  \sum_{j=1}^{N}p_{\theta}(i,j) \log_{2}(p_{\theta}(i,j)) -2\log_{e}(N) \nonumber \\
\widetilde{\mathrm{E}} &= \log_{e} (2) \cdot  \textrm{E} -2\log_{e}(N)
\end{align}
